\documentclass[12pt,aps,preprint,nofootinbib]{revtex4}

\usepackage{graphicx}
\usepackage{cancel}
\usepackage{amssymb}
\usepackage{textcomp}
\usepackage{amsmath}
\usepackage{bm}
\usepackage{times}
\usepackage{epsfig}
\usepackage{epstopdf}
\usepackage{color}
\usepackage{multirow}
\usepackage[colorlinks=true, pdfstartview=FitV, linkcolor=blue, citecolor=blue, urlcolor=blue]{hyperref}

\def\lsim{\mathrel{\raise.3ex\hbox{$<$\kern-.75em\lower1ex\hbox{$\sim$}}}}
\def\gsim{\mathrel{\raise.3ex\hbox{$>$\kern-.75em\lower1ex\hbox{$\sim$}}}}

\def\beq{\begin{equation}}
\def\eeq{\end{equation}}
\def\be{\begin{equation}}
\def\ee{\end{equation}}
\def\bea{\begin{eqnarray}}
\def\eea{\end{eqnarray}}

\def\bsg{b\to s\gamma}
\def\Br{\rm BR}

\def\ma{m_{A}}

\def\gev{\,{\rm GeV}}

\def\to{\rightarrow}
\def\met{\ensuremath{\not\!\!\!{E_{T}}}}

\newcommand{\minigraph}[5][0.25in]{\begin{minipage}{#2}\begin{center}\includegraphics[width=#2]{#5}\\\vspace{#3}\hspace{#1}{\footnotesize #4}\end{center}\end{minipage}}

\begin{document}

\preprint{~~PITT-PACC-1302}

\title{Non-Decoupling MSSM Higgs Sector and Light Superpartners}

%\vskip -1cm

\author{Tao Han$^{\bf a,e,f}$}
%\email{than@pitt.edu}

\author{Tong Li$^{\bf b}$}
%\email{than@pitt.edu}

\author{Shufang Su$^{\bf c}$}
%\author{Tong Li}
%\email{tli@udel.edu}

\author{Lian-Tao Wang$^{\bf d}$}
%\author{Tong Li}
%\email{tli@udel.edu}

\affiliation{
$^{\bf a}$  Pittsburgh Particle physics, Astrophysics, and Cosmology Center, Department of Physics and Astronomy, University of Pittsburgh, 3941 O'Hara St., Pittsburgh, PA 15260, USA\\
$^{\bf b}$  ARC Centre of Excellence for Particle Physics at the Terascale,
 School of Physics, Monash University,
 Melbourne, Victoria 3800, Australia \\
$^{\bf c}$
Department of Physics, University of Arizona, P.O.Box 210081, Tucson, AZ 85721, USA\\
$^{\bf d}$
Department of Physics, Enrico Fermi Institute, and Kavli Institute for Cosmological Physics,  University of Chicago, Chicago, IL 60637-1434, USA\\
$^{\bf e}$
Center for High Energy Physics,
%Department of Physics,
Tsinghua University, Beijing 100084, China\\
$^{\bf f}$
Korea Institute for Advanced Study (KIAS),
%School of Physics, 207-43 Cheongryangri-dong, Dongdaemun-gu
Seoul 130-012, Korea
}

\begin{abstract}
In the ``non-decoupling'' region of the Higgs sector in MSSM, the heavier CP-even Higgs boson ($H^0$) is Standard-Model-like and close to the charged Higgs bosons ($H^{\pm}$) in mass, while other neutral Higgs bosons ($h^0,\ A^0$) are lighter and near the $Z$ mass. This scenario is consistent with the current Higgs search limits, although the improved sensitivity for a light charged Higgs boson search $t\to H^{+}b$ may result in certain degree of tension. We demonstrate that  it can pass the stringent flavor constraints, provided there are other light SUSY particles to contribute in the loop induced processes. In turn, the non-decoupling Higgs sector implies the existence of light (left-handed) stop, sbottom and Wino-like gauginos, with mass all below 250 GeV.
These light super-partners can still escape the current SUSY searches at the LHC.  Dedicated searches for soft decay products should be devised for the LHC experiments to improve the searching sensitivity. The ILC would be able to cover the full spectrum region.
The solutions for the viable SUSY parameters result from subtle cancellations and are often missed by the generic multiple dimensional scans, highlighting the importance of theoretical guidance in search for such special cases.

\end{abstract}

\maketitle

\section{Introduction}

The milestone discovery of the Higgs boson in the LHC experiments \cite{:2012gk} has not only established the Standard Model (SM) as the correct effective theory to describe Nature up to the weak scale, but also opened a window to new physics associated with the Higgs sector.

One of the best motivated theories beyond the SM is the weak scale supersymmetry (SUSY).
In the framework of the Minimal Supersymmetric Standard Model (MSSM), the Higgs sector  \cite{Gunion:1989we,Djouadi:2005gj} has been extensively studied in the light of the recent Higgs discovery \cite{Carena:2011aa,Christensen:2012ei}.
It was elaborated \cite{Christensen:2012ei} that, for a Higgs boson of a mass $m_{h} \approx 126$ GeV, requiring the SM-like cross sections for $gg\to h \to \gamma\gamma,\ WW$,
the MSSM Higgs parameters split into two distinct regions:
\begin{itemize}
\item[(a)]  $m_A\lesssim130$~GeV:
 the ``non-decoupling" region \cite{Haber:1994mt}. In this region, the light CP-even Higgs $h^0$ and the CP-odd state $A^0$ are nearly mass degenerate and close to $m_{Z}$,
 while the heavy CP-even state $H^0$ is close to 126~GeV and the charged state $H^\pm$ is slightly heavier.
 \item[(b)] $\ma \gsim 400$ GeV: the ``decoupling'' region \cite{Haber:1994mt}. In this region, the light CP-even Higgs $h^0$ has a mass around 126~GeV, while all the other Higgs bosons are nearly degenerate with $\ma$ \cite{Haber:1995be}.
\end{itemize}
The non-decoupling scenario could be of immediate relevance for the LHC phenomenology:
If the non-SM Higgs bosons are all light, they may be accessible at the LHC even with the existing data~\cite{Christensen:2012si}.
In particular, the processes of Higgs pair production,
\begin{equation}
pp\to H^{\pm} A^{0},\ H^{+}H^{-},
\label{gauge}
\end{equation}
are via pure electroweak gauge interactions and are independent of the MSSM parameters except for their masses, in contrast to the gluon fusion and vector boson associated production processes.
Additionally, there may be sizable contributions from the processes
\begin{equation}
pp\to H^{\pm} h^{0},\ A^{0} h^{0},
\label{mssm}
\end{equation}
in the low-mass non-decoupling region which, on the other hand, do depend on the MSSM parameters and, thus, may be used to study the model.
We also note that the latest results from ATLAS charged Higgs searches in the $\tau+$ jets channel \cite{ATLAS_Hpm} have reached strong constraints on the branching fraction of $t \rightarrow b H^\pm$,
 which already resulted in certain degree of tension to this scenario.

However, it is well known that the $b$-quark rare decays via the flavor-changing neutral currents (FCNC) put very stringent constraints on the light Higgs and SUSY states \cite{Misiak:2006zs,bsmumuSM}.
It is thus not obvious if the non-decoupling region with light Higgs bosons would still be viable with respect to the stringent constraints from the flavor sector.
Because of the theoretical and observational importance, we set to explore this question in detail in the current work.
We not only scan over a broad range of SUSY parameter space, but also explore the detailed structure seeking for subtle cancellations.
Although the data from $B$ rare decays put significant bounds on the Higgs and gaugino sector, we find that this scenario can pass the stringent flavor constraints, provided there are other light SUSY particles to contribute in the loop induced processes.
The subtle cancellation among the contributing diagrams implies the existence of
a light (left-handed) stop with a mass typically in the range of 100 GeV$-$190 GeV,
a light (left-handed) sbottom of 160 GeV$-$250 GeV,
and light Wino-like gauginos of 100 GeV$-$250 GeV.
In addition, a Bino-like LSP is strongly favored due to the dark matter consideration in the MSSM  \cite{Han:2013gba}.
Taking into account the current stop and sbottom direct search limits from colliders, we find that there are two practical scenarios still viable to give consistent solutions:
$$
{\rm Case\ A:}\
M_1< M_2 < m_{\tilde{t}_1} < m_{\tilde{b}_1},
\ \  {\rm and}\ \
{\rm Case\ B:}\
M_1  < m_{\tilde{t}_1} < M_2 < m_{\tilde{b}_1} .
$$

Our findings are of both academic and practical significance: On the one hand, the solutions obtained   rely on cancellations that could be easily missed by generic scanning \cite{Christensen:2012ei,Conley:2010du}. The lesson for us is not to draw conclusions from simple generic features, since Nature might well be more subtle than naively expected.
On the other hand, the solutions can be viewed as a sharp prediction.
While the dominant stop decay is $\tilde{t}_1 \rightarrow bW^{(*)}\tilde{\chi}_1^0$ or $\tilde{t}_1\to c\tilde{\chi}_1^0$,
 the sbottom mostly decays via
$\tilde{b}_1 \rightarrow b \tilde{\chi}_2^0$ with $\tilde{\chi}_2^0$ decay subsequently to $Z^{(*)} \tilde{\chi}_1^0$, $h^{0(*)} \tilde{\chi}_1^0$, etc.    The study of such a light stop and sbottom, as well as $\tilde{\chi}_2^0$ at the LHC poses special interest for the Higgs non-decoupling region, which calls for further theoretical and dedicated experimental investigations.
An $e^{+}e^{-}$ International Linear Collider (ILC), on the other hand, would be able to cover the full spectrum region, once it crosses over the mass threshold of the SUSY partners.

The rest of the paper is organized as follows. In Sec.~\ref{sec:MSSMH},
we summarize the SUSY contribution to rate $B$ decays, in particular, to $\bsg$.
In Sec.~\ref{ScanS},
we recall the Higgs sector of the MSSM in the non-decoupling region and present the parameter choices for our scan.  we also present our results with the scanning subject to the stringent flavor constraints.
We discuss the consequences of the scan results for the LHC searches of stop, sbottom and neutralinos/charginos and also comment on the potential benefits for their studies at the ILC in Sec.~\ref{Disc}. We summarize our results in Sec.~\ref{Concl}. We collect the formulae of loop functions for flavor observables in the Appendix.

%%%%%%%%%%%%%%%%%%%%%%%%%%%%%%%%%%%%%%%%%%%%%%%%

%\section{Non-Decoupling MSSM Higgs sector and Flavor constraints}
\section{Non-Decoupling MSSM Higgs sector confronting $B$ rare decays}
\label{sec:MSSMH}

The non-decoupling MSSM Higgs scenario has been mapped out in some recent studies \cite{Christensen:2012ei, Heinemeyer,Hagiwara,Ke:2012zq,Wagner}. This scenario necessarily has a light charged Higgs. However, its contributions to flavor changing neutral current processes \cite{Grinstein:1987pu}, such as $\bsg$,  are too large to be consistent with the measurements. Additional contributions can also arise in MSSM, mediated by loops containing super-partners. It is possible that such SUSY contributions \cite{Bertolini:1987pk} can partially cancel the dangerous charged Higgs contribution, rendering this scenario phenomenologically viable. Short of a symmetry, such a cancellation is indeed finely tuned. Nevertheless, given the central roles of supersymmetry and electroweak symmetry breaking in our speculations on new physics beyond the Standard Model, and the lack of knowledge of flavor physics in SUSY theories, it is important to leave no stone unturned in covering the SUSY parameter space.

In general, there are many possible new contributions to FCNC in supersymmetry. For example, squark mass matrices can have off-diagonal entries. In this case, gluino-squark diagrams with such flavor violating couplings give some of the largest new contributions to FCNC. As a result, many of the off-diagonal entries in the squark mass matrices are strongly constrained. This is a well known example of the SUSY flavor problem.  While interesting and acceptable effects from such couplings are still possible, it needs to be addressed in a comprehensive framework. In this work, we follow an alternative approach which does not rely on new flavor violation from SUSY breaking, assuming they have been forbidden or strongly suppressed. In this case, supersymmetric charged-current couplings still give new contributions to FCNC processes. We note that this is the simplest realization of the Minimal Flavor Violation scenario  \cite{MFV}. In this case, satisfying constraints from $B$ rare decay measurements requires the stop and sbottom to be light. Hence, this scenario necessarily predicts improved naturalness of the electroweak symmetry breaking.
We now provide a qualitative discussion of the necessary cancellation.

The $\Delta F=1$ effective Hamiltonian relevant for  the $b\to s\gamma$ transition is \cite{Inami:1980fz}
\begin{eqnarray}
H_{\rm eff}\supset -{4G_F\over \sqrt{2}}V_{ts}^\ast V_{tb}\left(C_7 O_7+C_8 O_8\right),
\end{eqnarray}
where the magnetic and chromomagnetic operators  are
\begin{eqnarray}
O_7 = {e\over 16\pi^2}m_b(\bar{s}_L\sigma^{\mu\nu}b_R)F_{\mu\nu},\quad
O_8 = {g_s\over 16\pi^2}m_b(\bar{s}_L\sigma^{\mu\nu}T^a b_R)G_{\mu\nu}^a.
\label{eq:bsg_opt}
\end{eqnarray}

The Standard Model contribution to $\bsg$ is dominated by the $W$-top  loop  \cite{Inami:1980fz}. In our scenario, the most model-independent new contribution comes from the $H^{\pm}$-top  loop \cite{Grinstein:1987pu}.
Their contributions to $C_7$ and $C_8$ are
\begin{eqnarray}
C_{7,8}^{H^\pm}\simeq f_{7,8}\left({m_t^2\over m_{H^\pm}^2}\right),
\end{eqnarray}
where the loop functions $f_{7,8}$ are given in Eq.~(\ref{eq:f78}) in the Appendix. It is positive and the same as the SM contribution.
The charged Higgs contribution is proportional to the top Yukawa coupling $y_t^2$,  independent of $\tan \beta$ at the leading order.
For the parameter region in the non-decoupling scenario, this contribution leads to a BR$(\bsg)$   about factor of two larger than the experimental observations for $m_{H^\pm}$ around 100 GeV.  Therefore, taking into account a $10\%$ accuracy of the current measurements, about $5\%$ fine-tuning is needed from other SUSY contributions to $\bsg$ in order to cancel the large positive contribution from the charged Higgs sector.

\begin{figure}[tb]
\centering
\minigraph{5cm}{-0.15in}{(a)}{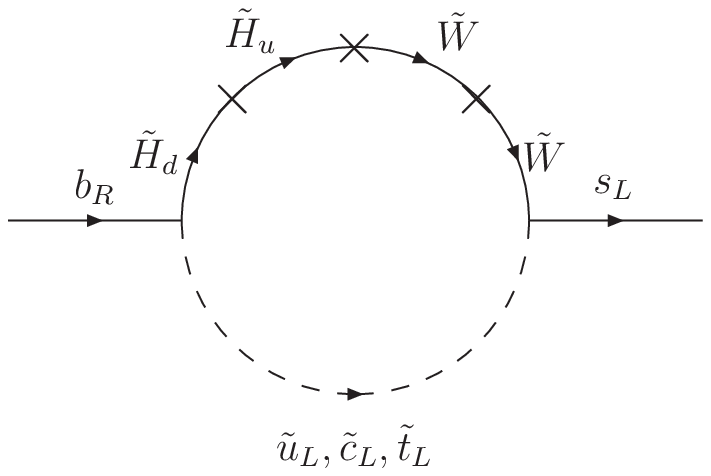}
%\includegraphics[scale=1,width=5cm]{../plots/d.pdf}
%\hfill
\minigraph{5cm}{-0.15in}{(b)}{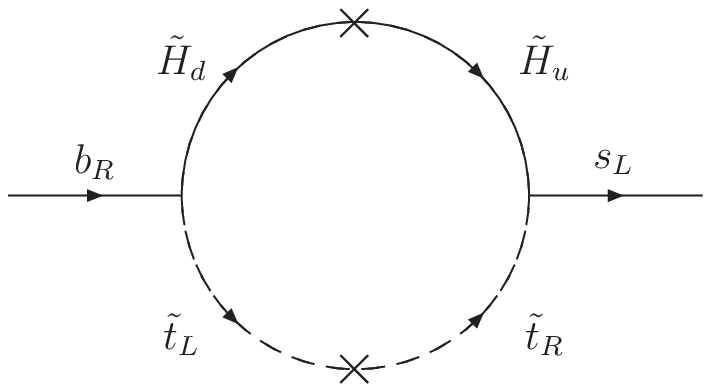}\\
\minigraph{5cm}{-0.15in}{(c)}{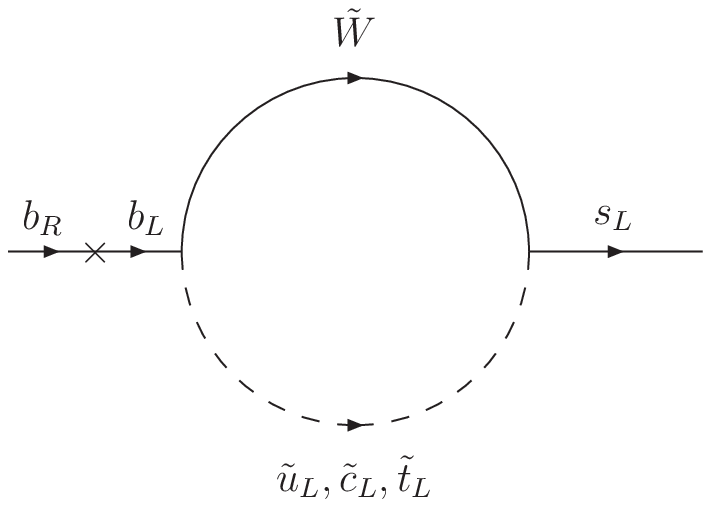}
\minigraph{5cm}{-0.15in}{(d)}{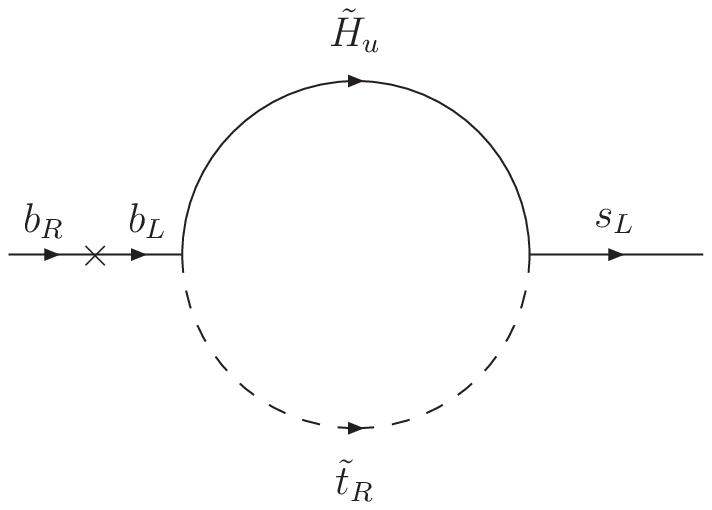}
\minigraph{5cm}{-0.15in}{(e)}{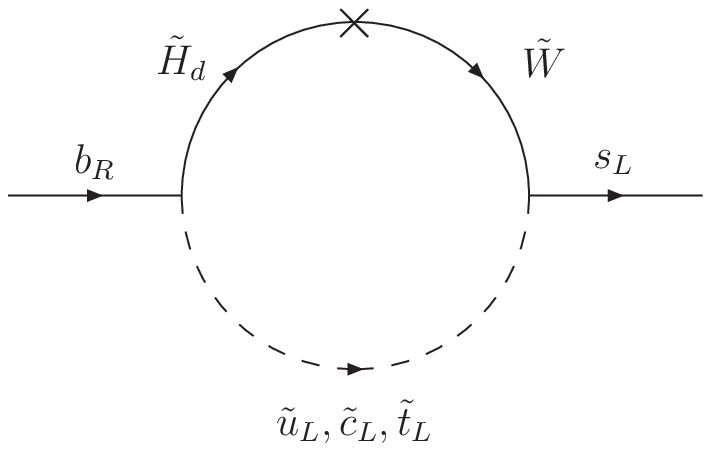}
\caption{Dominant SUSY contributions to $\bsg$.}
\label{fig:bsg_SUSY}
\end{figure}

While the charged Higgs loops always contribute constructively to the SM, the SUSY chargino loops could contribute with either sign, depending on the SUSY parameters. As usual, it is more intuitive to discuss the parameter-dependence by studying the SUSY loop contributions in the gauge eigenstate basis of $\tilde{H}_u^+, \tilde{H}_d^-, \tilde{W}^\pm, \tilde{t}_L, \tilde{t}_R$.
The important SUSY contributions are shown in Fig.~\ref{fig:bsg_SUSY}.
Note a chirality flip is necessary for the diagrams to contribution to  $\bsg$ magnetic operator as in Eq.~(\ref{eq:bsg_opt}), which occurs via the insertion of an odd number of  fermion masses, including $m_{b}$, $m_t$, and/or super-partner masses $M_2$, $\mu$ and the Higgsino-Wino mixing. Depending on the nature of the couplings, the left-right mixing in the squark sector may be required at the same time.

In order to cancel the charged Higgs contribution, we expect at least some of the super-partners would have to be very light, with masses similar to that of the charged Higgs.\footnote{One possible alternative is to have enhanced large couplings. However, within our scenario, this effect is limited. For example, as we discussed later, $\tan \beta$ is strongly constrained by collider searches and cannot be very large, and it does not negate the requirement of very light super-partners.}  The two largest contributions come from diagrams Fig.~\ref{fig:bsg_SUSY}(a) and Fig.~\ref{fig:bsg_SUSY}(b), both are enhanced by $\tan\beta$:
\begin{eqnarray}
C_{7,8}^{(a)}&\simeq& 2M_W^2 \mu M_2\tan\beta\left[{1\over M_{\tilde{Q}_3}^4}g^{(a)}_{7,8}\left({M_2^2\over M_{\tilde{Q}_3}^2},{\mu^2\over M_{\tilde{Q}_3}^2}\right)-{1\over M_{\tilde{q}}^4}g^{(a)}_{7,8}\left({M_2^2\over M_{\tilde{q}}^2},{\mu^2\over M_{\tilde{q}}^2}\right)\right],
\label{eq:bsg_a}
\\
C_{7,8}^{(b)}&\simeq&
-{m_t^2 \mu A_t\over 2M_{\tilde{Q}_3}^4}\tan\beta \ g^{(b)}_{7,8}\left({\mu^2\over M_{\tilde{Q}_3}^2},{\mu^2\over M_{\tilde{t}_R}^2}\right),
\label{eq:bsg_b}
\end{eqnarray}
where the loop functions $g_{7,8}^{(a), (b)}$ are given in Eqs.~(\ref{eq:g7a})-(\ref{eq:g8b}) in the Appendix.
$M_{\tilde{Q}_3}$ ($M_{\tilde{t}_R}$) is the mass parameter for the left(right)-handed stop and $M_{\tilde{q}}$ is the common squark mass parameter for the first two generation squarks.  Note that in the case of $M_{\tilde{Q}_3}^2=M_{\tilde{q}}^2$, the SUSY contribution is small due to super-GIM suppression. However, we are working in the limit of $M_{\tilde{q}} \gg M_{\tilde{Q}_3}$ and the super-GIM cancellation is not relevant  for either process in Fig.~\ref{fig:bsg_SUSY}(a)  or those discussed later.

Figs.~\ref{fig:bsg_SUSY}(a) shows the loop contribution from $\tilde{H}^\pm/\tilde{W}^\pm$-$\tilde{t}_L$, with the chirality flip via
fermion mass insertions indicated.  Even with moderate values of $\tan \beta \sim 5 - 10$, the enhancement proportional to $\tan\beta$ is still significant.  The contribution to $\bsg$ is negative for $\mu M_2>0$, and its value can be as large as factor of two  of the charged Higgs contributions in the non-decoupling region.  Therefore,  this diagram  is very important in canceling the charged Higgs contribution if $M_2$ and  $M_{\tilde{Q}_3}$ are not much larger than $m_{H^\pm}$.

Fig.~\ref{fig:bsg_SUSY}(b) shows the contribution of $\tilde{H}$-$\tilde{t}_L / \tilde{t}_R$ loop, with the chirality flip achieved via Higgsino mass insertion. A stop left-right mixing, $ m_t {A}_t$, is also required.
The contribution of this diagram  can be comparable in size to the charged Higgs contribution. At the same time, it is still somewhat smaller than the contribution in Fig.~\ref{fig:bsg_SUSY}(a), due to the heavy right-handed stop in the loop.\footnote{Note that $M_{\tilde{t}_R}$ has to be heavy in order to accommodate the sizable SM-like Higgs mass while keeping $M_{\tilde{Q}_3}$ small.}
It is positive for $\mu A_t>0$ and flips sign for $\mu A_t<0$. There is another term in the stop left-right mixing, which gives a contribution proportional to $|\mu|^2$ that is not $\tan \beta $ enhanced. Such term is only important for $\mu /\tan\beta > A_t$, which is not realized in the parameter region that we are considering.

Fig.~\ref{fig:bsg_SUSY}(c) shows the $\tilde{W}^\pm$-$\tilde{t}_L$ loop, with the chirality flip in bottom quark.  The contribution from this diagram depends on $M_2$ and $M_{\tilde{Q}_3}$:
\begin{eqnarray}
C_{7,8}^{(c)}\simeq 2M_W^2\left[{1\over M_{\tilde{Q}_3}^2}g^{(c)}_{7,8}\left({M_2^2\over M_{\tilde{Q}_3}^2}\right)-{1\over M_{\tilde{q}}^2}g^{(c)}_{7,8}\left({M_2^2\over M_{\tilde{q}}^2}\right)\right],
\end{eqnarray}
where the loop functions $g_{7,8}^{(c)}$ are given in Eq.~(\ref{eq:g78c}) in the Appendix.  We find that, even for $M_{\tilde{Q}_3} \sim m_{H^{\pm}}$, it is numerically much smaller than the charged Higgs contribution, as it is suppressed by at least a relative factor $g_2^2/y_t^2$. Therefore, although it is of the opposite sign, it cannot play an important role in satisfying the $\bsg$ constraint.

Fig.~\ref{fig:bsg_SUSY}(d) shows the $\tilde{H}_u$-$\tilde{t}_R$ loop, with again the chirality flip in bottom quark propagator.  The contribution from this diagram depends on $|\mu|$ and $M_{\tilde{t}_R}$:
\begin{eqnarray}
C_{7,8}^{(d)}\simeq
{m_t^2\over M_{\tilde{t}_R}^2}g^{(c)}_{7,8}\left({\mu^2\over M_{\tilde{t}_R}^2}\right).
\end{eqnarray}
Given the viable range of large $\mu$
and $M_{\tilde{t}_R}$, contribution from Fig.~\ref{fig:bsg_SUSY}(d) is always negative and relatively small. A similar diagram with $\tilde{H}_d$-$\tilde{t}_L$  is also small given the extra bottom Yukawa suppression.

Fig.~\ref{fig:bsg_SUSY}(e) shows the $\tilde{H}_d/\tilde{W}-\tilde{t}_L$ loop with the
  loop contributions being
\begin{eqnarray}
C_{7,8}^{(e)}&\simeq& 2M_W^2\left[{1\over M_{\tilde{Q}_3}^2}g^{(e)}_{7,8}\left({M_2^2\over M_{\tilde{Q}_3}^2},{\mu^2\over M_{\tilde{Q}_3}^2}\right)-{1\over M_{\tilde{q}}^2}g^{(e)}_{7,8}\left({M_2^2\over M_{\tilde{q}}^2},{\mu^2\over M_{\tilde{q}}^2}\right) \right],
\label{eq:bsg_e}
\end{eqnarray}
where the loop functions $g_{7,8}^{(e)}$ are given in Eqs.~(\ref{eq:g7e})-(\ref{eq:g8e}) in the Appendix. The contribution from Fig.~\ref{fig:bsg_SUSY}(e) is also small and negative.

We reiterate that the non-decoupling region favors $\mu A_t>0$. This is because a large positive radiative correction to the bottom Yukawa, $\Delta m_b$, is needed to suppress $H^0\to b\bar{b}$ so as to enhance $gg\to H^0\to \gamma\gamma$~\cite{Christensen:2012ei}. Since we choose $M_2,\ \mu >0$, which is favored from the muon $g-2$ consideration, $A_t$ has to be positive. Fig.~\ref{fig:bsg_SUSY}(b) is the only diagram that depends on $A_t$, which gives a positive contribution to $b\to s\gamma$ for $\mu A_t>0$.  The contribution from Fig.~\ref{fig:bsg_SUSY}(a), however, is large and negative
when  lowering $M_2$ as well as $M_{\tilde{Q}_3}$.  Therefore, the large positive contributions from both the charged Higgs and
Fig.~\ref{fig:bsg_SUSY}(b) can be cancelled by contributions from Fig.~\ref{fig:bsg_SUSY}(a), with about 5\% fine-tuning.

$B_s\to \mu^+ \mu^-$ does not impose additional very stringent constraint on our scenario. It is not nearly as precisely measured as the $b\to s \gamma$. It is well known that MSSM Higgs amplitude can be proportional to $\tan^3 \beta$~\cite{bsmumuSM}, and hence very important at large $\tan \beta$, which are disfavored for the non-decoupling scenarios. In fact, the limits on $\tan \beta$ from direct collider searches seems to be always stronger than the one from $B_s\to \mu^+ \mu^-$~\cite{Altmannshofer:2012ks}.

We see that satisfying the flavor constraints in the non-decoupling scenario naturally leads to the presence of light third generation squarks, in particular, $\tilde{t}_L$ and $\tilde{b}_L$.  Given the close connection of the third generation squarks to the Higgs sector, there have been extensive collider studies on the direct collider searches for the stops and sbottoms at the Tevatron and the LHC. We will discuss the implications of collider searches on our scenario in Section IV.

\section{SUSY Parameter region and experimental bounds}
\label{ScanS}

To explore the parameter space which is consistent with the current observation including the flavor constraints, we follow the procedure as in Ref.~\cite{Christensen:2012ei}, and perform a comprehensive scan over the MSSM parameter space
\begin{eqnarray}
\nonumber
& 3 < \tan\beta < 55, \quad
 50\ {\rm GeV}< m_A < 500\ {\rm GeV}, \quad 100\gev <  \mu < 1000\ {\rm GeV}, &\\
&   100\gev < M_{\tilde{t}_R}, M_{\tilde{Q}_3} < 2000\ {\rm GeV}, \quad
    {-4000\ \gev}< A_t < 4000\  {\rm GeV}. &
\label{eq:para}
\end{eqnarray}
To study the non-decoupling region, we focus on the reduced low $\ma$ range with positive $A_t$:
\begin{equation}
 95\ {\rm GeV}< m_A < 130\ {\rm GeV}  , \quad 0<A_t<4000\ {\rm GeV}.
 \label{eq:ma}
\end{equation}
The $SU(2)_L$ gaugino mass parameter has significant impact on the flavor sector in the non-decoupling region and we scan it over a low-value range
\begin{eqnarray}
100 \ {\rm GeV}< M_2 < 300\ {\rm GeV}.
\end{eqnarray}
Other SUSY soft masses, which are less relevant to our consideration, are all fixed to be 3 TeV.

%%%%%%%%%%%%%%%%%%%%%%%%%%

\subsection{Constraints from the Higgs signal  and exotic Higgs searches }

We perform our scan by using the FeynHiggs 2.9.4 package~\cite{Degrassi:2002fi,Heinemeyer:1998np,Frank:2006yh,Heinemeyer:1998yj} to calculate the mass spectrum, couplings and other SUSY parameters. We use HiggsBound 3.8.1 \cite{Bechtle:2008jh} to check the exclusion constraints from LEP2~\cite{LEP2H}, the Tevatron~\cite{CDFD0} and the LHC~\cite{ATLASrr,ATLASww,ATLASzz,CMSrr,CMSww,CMSzz,ATLASvhbb,ATLAStthbb,CMSvhbb,CMStthbb,ATLAStata,ATLAStataMSSM,CMSmumuMSSM,CMSbbMSSM,ATLASHpmta,CMSHpmta,ATLASHpmcs}. The latest LHC Higgs search results \cite{moriondATLAS,moriondCMS,mssmtau,ATLAS_Hpm} are implemented by hand.  We generate a large random data sample that passes these constraints.
We require that the heavy CP-even Higgs boson is  SM-like and satisfies the following properties
\begin{eqnarray}
&&H^0 \ {\rm in \ the \ mass \ range \ of} \ 124 \ {\rm GeV} - 128 \ {\rm GeV},\\
&&  \sigma \times {\Br} (gg\to H^0   \to \gamma\gamma)_{\rm MSSM}\geq 90\% (\sigma\times {\Br})_{\rm SM}.
\end{eqnarray}

We note that the recent search for MSSM neutral Higgs in the $\tau\tau$ mode by CMS apparently excludes
$\tan\beta>5$ for $M_A < 250 $ GeV~\cite{mssmtau}.
However, that exclusion is made under particular assumptions. First of all, the limit is obtained in the so-called $m_h^{\rm max}$ scenario, with $M_{\rm SUSY}$ assumed to be large. Recently, Ref.~\cite{Djouadi} has argued that this limit can be extended beyond the $m_h^{\rm max}$ scenario. Similar SUSY scenarios were studied in Ref.~\cite{Arganda:2012qp}. However, in our scenario, there are additional light squarks and gauginos, which can induce a significant shift of Higgs couplings to the SM quarks.
For example, studies in Refs.~\cite{Heinemeyer,Wagner} showed that $\tan \beta$ as large as $8-9$ are still allowed.
Given these unsettled arguments, we set out a scanning procedure over a broad range in the SUSY parameter space, with results for imposing the CMS $\tau\tau$ exclusion limits presented in Sec.~\ref{sec:results}.

%will not attempt to recast the CMS search, and we will consider a broader range of $\tan \beta$.

The latest results from ATLAS charged Higgs searches in the $\tau+$ jets channel \cite{ATLAS_Hpm} impose strong constraints on the branching fraction of $t \rightarrow b H^\pm$ for $m_{H^\pm} < m_t$, with the 95\% C.L. upper limits  in the range of 2.1\% to 0.24\% for $m_{H^\pm}$ between 90 GeV to 160 GeV.
% (corresponding to $M_A$ in the range of 95 to 130 GeV).
We imposed this charged Higgs search limit in our parameter scan at $3\sigma$ level of their errors.

\subsection{Constraints from $b$ Rare Decays}

The experimental measurements considered in this study include $\bsg$ \cite{Amhis:2012bh} and
the LHCb report on $B_s\to \mu^+ \mu^-$ \cite{Aaij:2012nna}. In our study, we use the following SM predictions  \cite{Misiak:2006zs,Misiak:2006ab,bsmumuSM}, and experimental limits,
\begin{eqnarray}
{\Br}(B_s \rightarrow X_s \gamma)_{\rm exp} = (3.43\pm 0.21)\times 10^{-4},&&
{\Br}(B_s \rightarrow X_s \gamma)_{\rm SM} = (3.15\pm 0.23)\times 10^{-4},~~~\quad \\
{\Br}(B_s\to \mu^+ \mu^-)_{\rm exp} = (3.2^{+1.5}_{-1.2})\times 10^{-9},&&
{\Br}(B_s\to \mu^+ \mu^-)_{\rm SM} = (3.23\pm 0.27)\times 10^{-9}.
\end{eqnarray}
BABAR also reported improved measurements of $B\to D \tau \nu_\tau$ which indicates a deviation from the SM expectation and is sensitive to new physics contributions in the form of a light charged Higgs boson at tree level. We take the observed excess as an upper limit~\cite{Lees:2012xj}
\begin{eqnarray}
&& {{ {\Br}(B\to D\tau\nu_\tau) } \over { {\Br}(B\to D\ell\nu_\ell)}} <0.44 , \quad
{{ {\Br}(B\to D\tau\nu_\tau)_{\rm SM}} \over { {\Br}(B\to D\ell\nu_\ell)_{\rm SM} }} =0.297\pm 0.017 , \\
&& {{ {\Br}(B\to D^{\ast}\tau\nu_\tau) } \over { {\Br}(B\to D^{\ast}\ell\nu_\ell)}} < 0.332, \ \
{{ {\Br}(B\to D^{\ast}\tau\nu_\tau)_{\rm SM}} \over { {\Br}(B\to D^{\ast}\ell\nu_\ell)_{\rm SM} }} = 0.252\pm 0.003.
\end{eqnarray}
In our numerical study, we use SuperIso 3.3~\cite{superiso} to evaluate the above flavor observables. We also check that the values of ${\Br}(B\to \tau\nu)$, $\Delta m_{B_d}$ and $\Delta m_{B_s}$ in the non-decoupling region are consistent with the recent Belle~\cite{btaunu} and LHCb~\cite{deltam} measurements, respectively. As discussed in the previous section, the $\bsg$ process gives the most stringent constraint on non-decoupling region with light charged Higgs. To achieve the SM-like measurement, in MSSM, the dominant charged Higgs-top loop contribution should be largely cancelled by other SUSY loops, in particular, the chargino-stop loop, which requires small values of $M_2$ and $M_{\tilde{Q}_3}$.

%%%%%%%%%%%%%%%%%%%%%%%%%

\begin{figure}[t]
\centering
\includegraphics[width=0.49\textwidth]{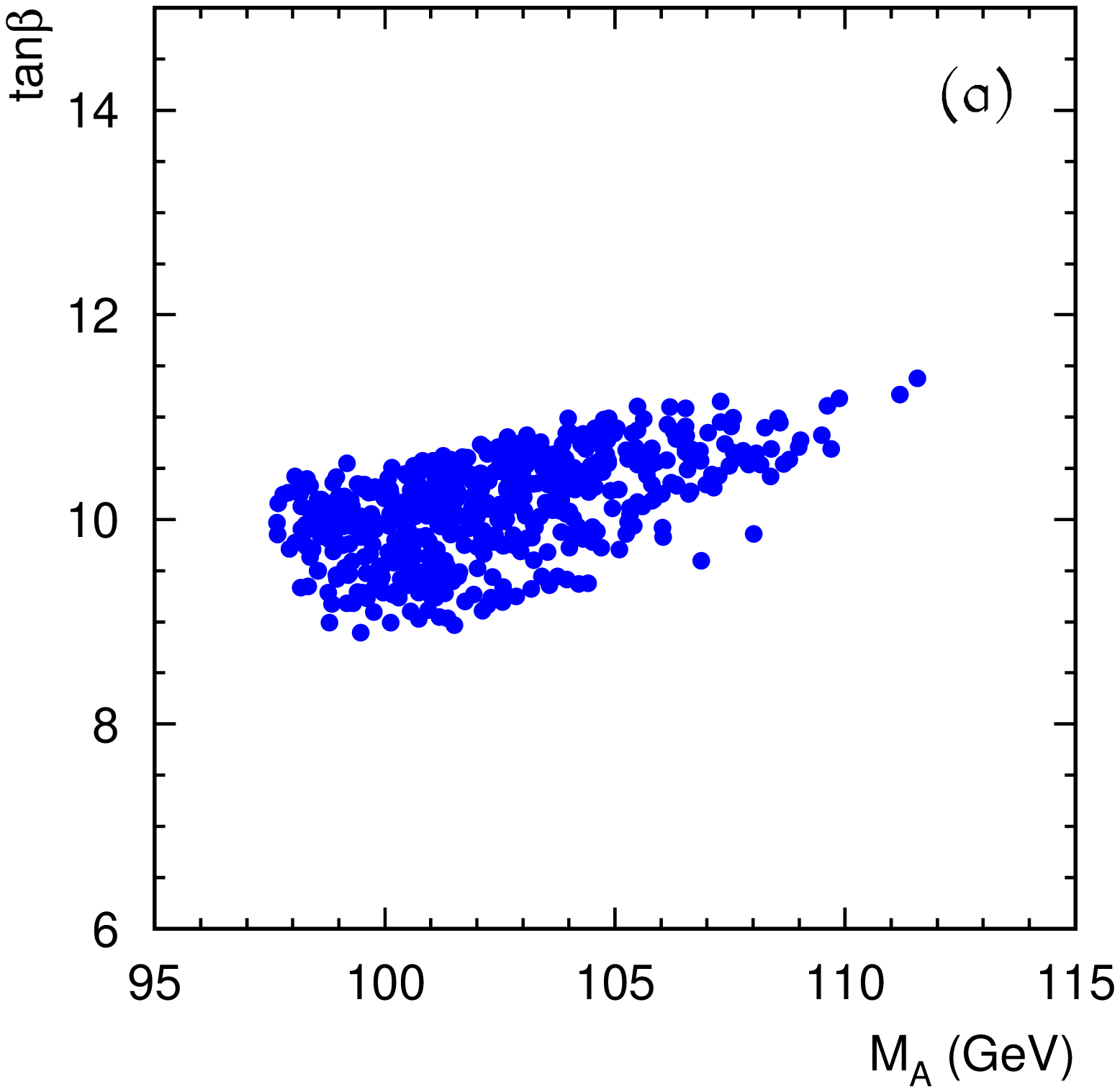}
\includegraphics[width=0.49\textwidth]{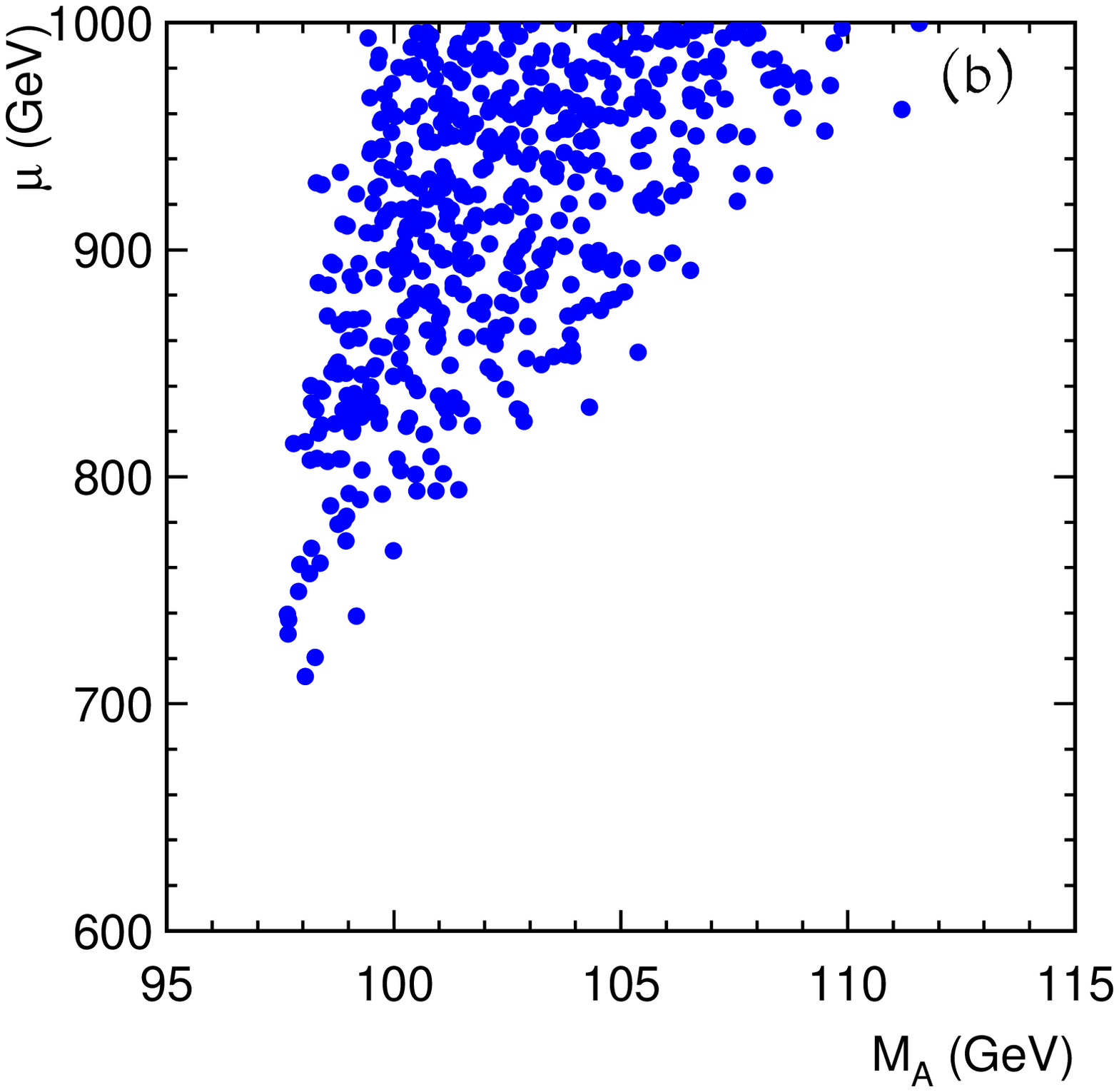}
\includegraphics[width=0.49\textwidth]{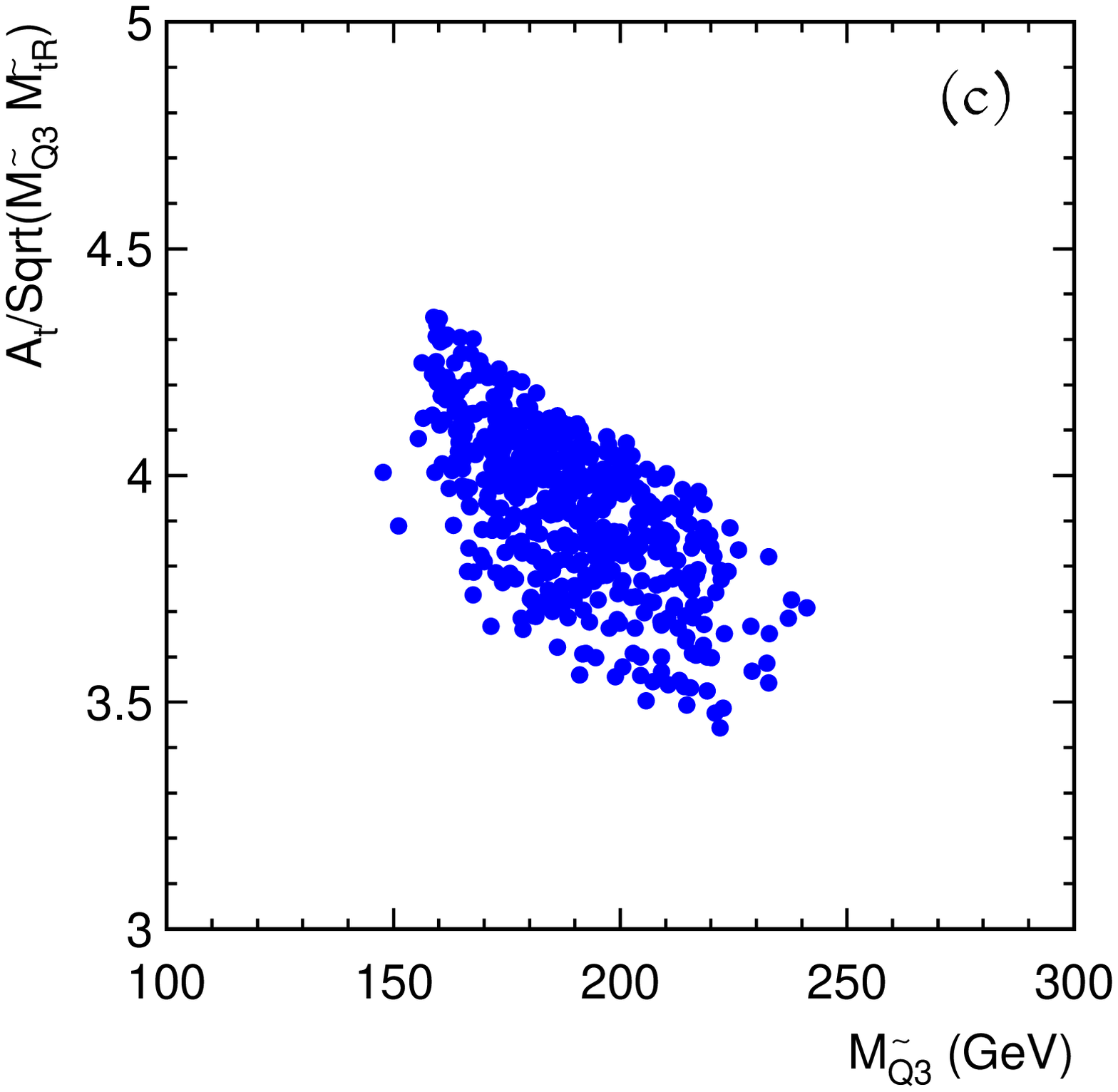}
\includegraphics[width=0.49\textwidth]{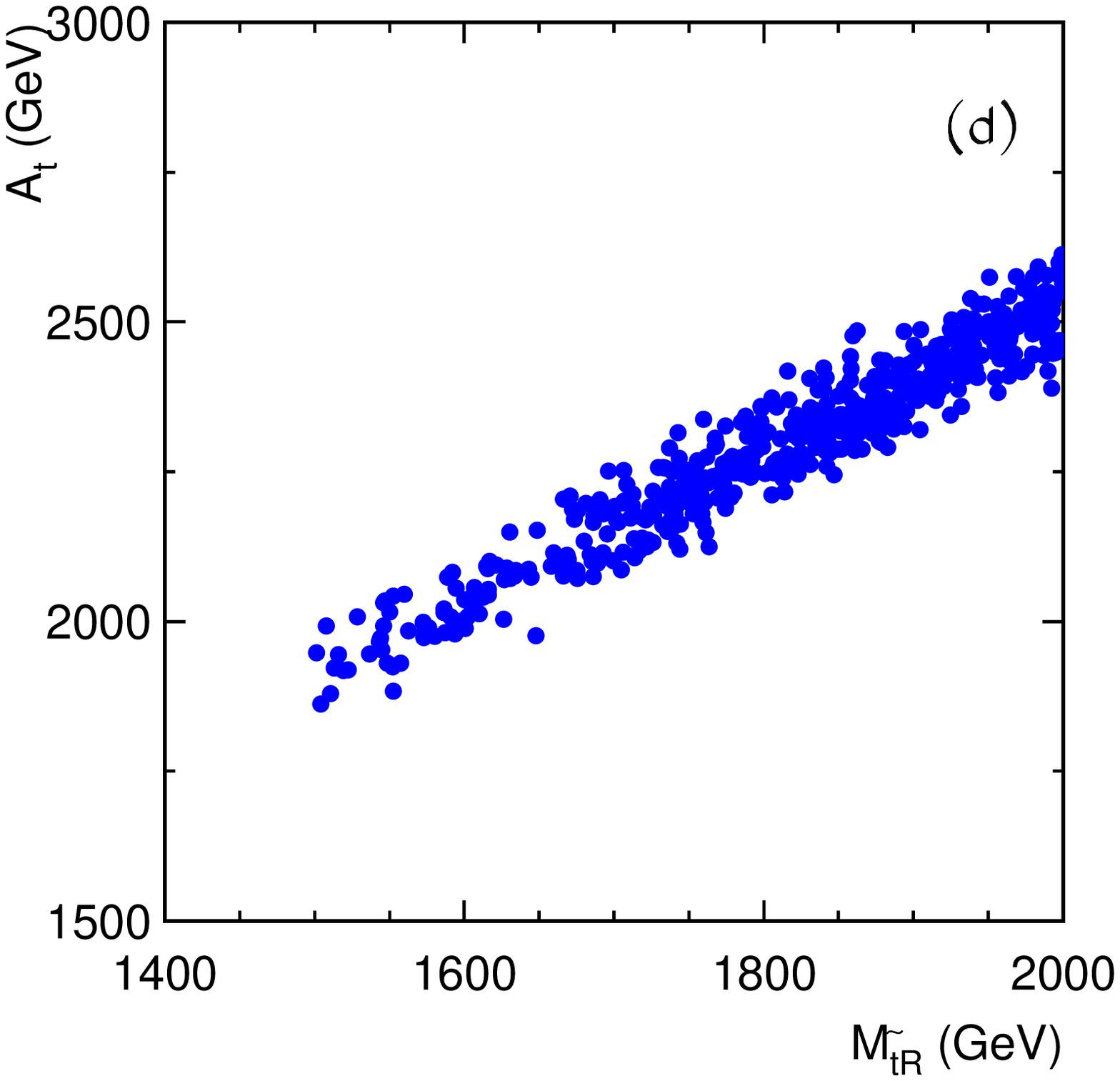}
\caption{Allowed regions for the Higgs sector parameters (a) $\tan\beta$ versus $m_A$, (b) $\mu$ versus $m_A$,  (c) $A_{t}/\sqrt{M_{\tilde{Q}_3} M_{\tilde{t}_R}}$ versus $M_{\tilde{Q}_3}$, and (d) $A_{t}$ versus $M_{\tilde{t}_R}$.
}
 \label{fig:MA}
\end{figure}

%%%%%%%%%%%%%%%%%%%%%%%%%

\subsection{Results for Allowed Region}
\label{sec:results}

Taking into account both the Higgs search results and the flavor constraints,
we first show the surviving points in Fig.~\ref{fig:MA} in the parameter space relevant for the MSSM Higgs sector:  (a) $\tan\beta$ versus $m_A$, (b) $\mu$ versus $m_A$, (c) $A_{t}/\sqrt{M_{\tilde{Q}_3} M_{\tilde{t}_R}}$ versus $M_{\tilde{Q}_3}$, and (d) $A_{t}$ versus $M_{\tilde{t}_R}$.  The range of $\tan\beta$, $m_A$, $\mu$ is found to sit in the MSSM non-decoupling region:
\begin{equation}
9 < \tan\beta < 11, \quad
 97\ {\rm GeV}< m_A < 113\ {\rm GeV}, \quad 700\gev <  \mu < 1000\ {\rm GeV} ,
 \end{equation}
and nearly maximal stop mixing:
 \begin{equation}
3.5 < A_t/\sqrt{M_{\tilde{Q}_3} M_{\tilde{t}_R}} < 4.5.
 \end{equation}
The mass parameter for $\tilde{t}_L$, $M_{\tilde{Q}_3}$, however, is constrained to be small to maximize the   negative contribution to $\bsg$ from $\tilde H / \tilde W$-$\tilde{t}_L$ loop [Fig.~\ref{fig:bsg_SUSY}(a)]  in order to cancel the positive contribution from a light charged Higgs.  Consequently, large  $M_{\tilde{t}_R}$ and $A_t$  are needed to raise the Higgs mass value to 124 $-$ 128 GeV and we find:
\begin{equation}
150\gev < M_{\tilde{Q}_3} < 240\ {\rm GeV},\
1500\gev < M_{\tilde{t}_R} < 2000\ {\rm GeV},\ {1800\ \gev}< A_t < 2600\  {\rm GeV}.
\end{equation}

\begin{figure}[t]
\centering
\includegraphics[width=0.49\textwidth]{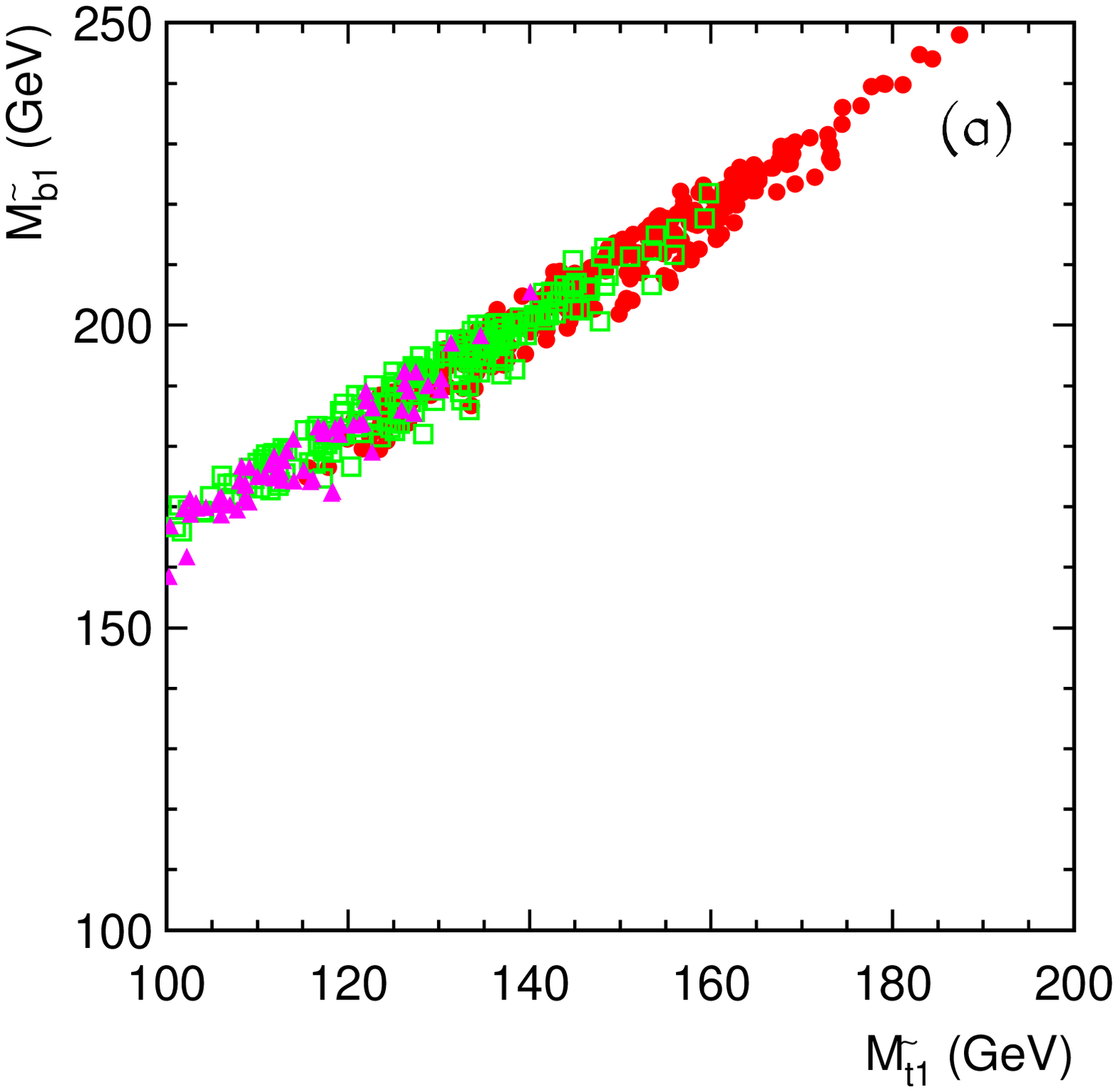}
\includegraphics[width=0.49\textwidth]{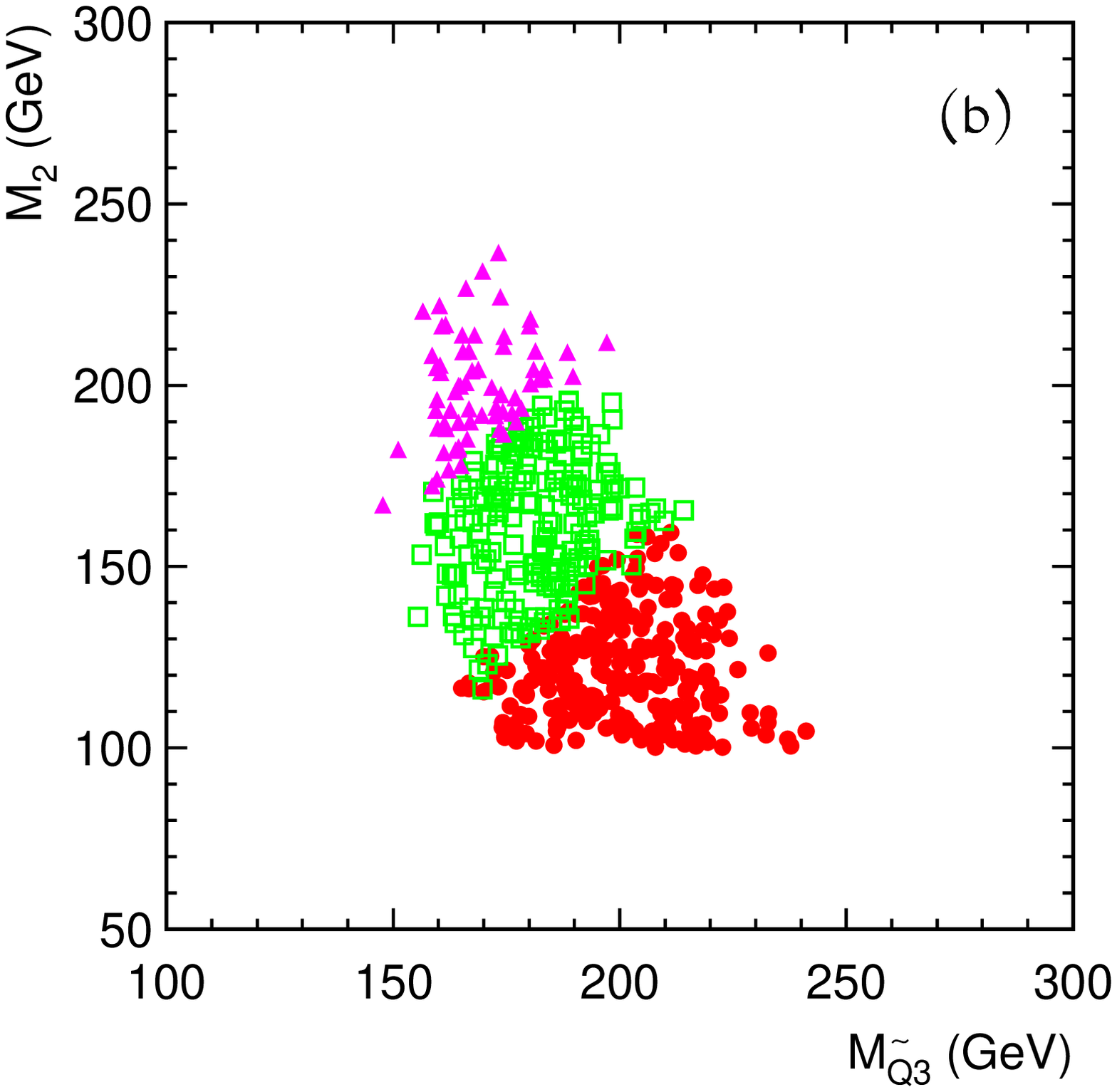}
\includegraphics[width=0.49\textwidth]{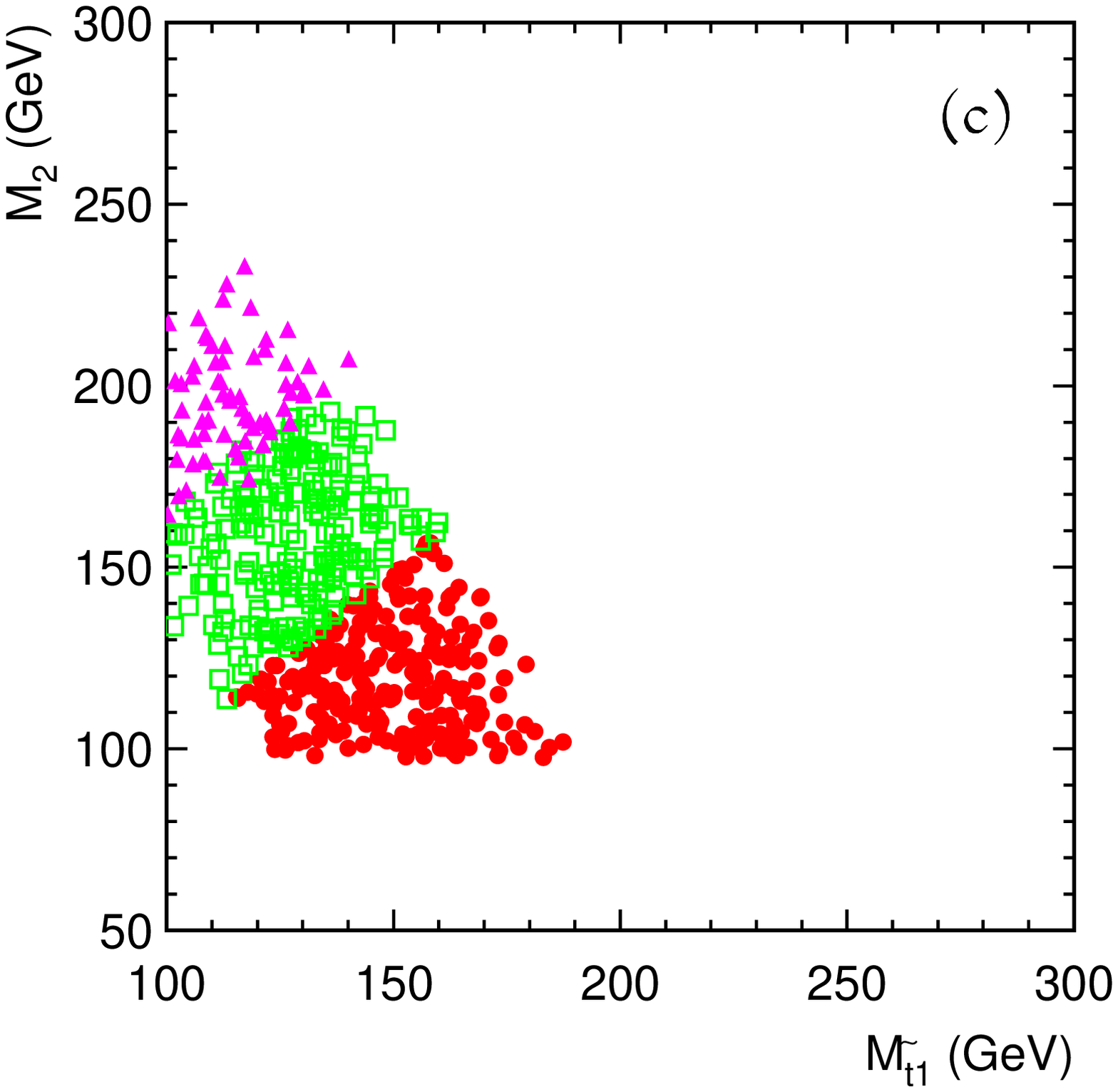}
\includegraphics[width=0.49\textwidth]{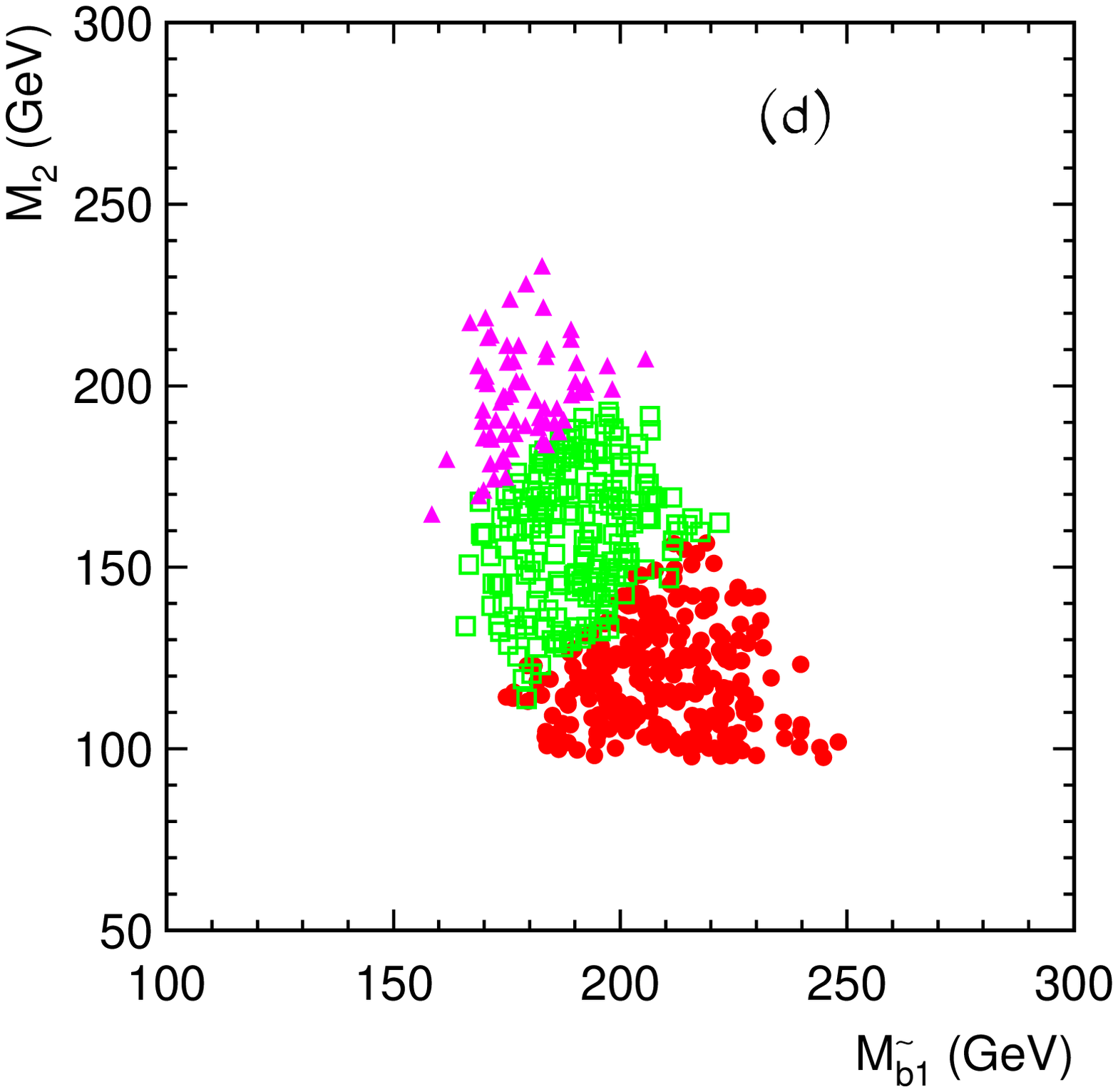}
\caption{Allowed regions for SUSY mass parameters (a) $m_{\tilde{b}_1}$ versus $m_{\tilde{t}_1}$,
(b) Wino mass $M_2$ versus  $M_{\tilde{Q}_3}$, (c)  $M_2$ versus $m_{\tilde{t}_1}$,  and (d) $M_2$ versus $m_{\tilde{b}_1}$.
The red circles denote the case with $M_{2} < m_{\tilde{t}_{1}} < m_{\tilde{b}_1}$, the green squares denote the case with $m_{\tilde{t}_{1}} < M_{2} < m_{\tilde{b}_1}$, and the purple triangles denote the case with $m_{\tilde{t}_{1}} < m_{\tilde{b}_1} < M_{2}$.
}
\label{fig:M2}
\end{figure}
The above results indicate the presence of  a light stop which is mostly $\tilde{t}_L$, as well as a light sbottom $\tilde{b}_L$, both of which are controlled by the same mass parameter $M_{\tilde{Q}_3}$.

In Figure \ref{fig:M2}, we show the allowed regions for SUSY mass parameters subject to various constraints under consideration. The three color codes indicate the relative value of $M_{2}$ with respect tot he squark masses, as specified in the figure caption.
Figure \ref{fig:M2}(a) shows $m_{\tilde{b}_1}$ versus $m_{\tilde{t}_1}$, from which one reads
\be
100~\gev < m_{\tilde{t}_1} < 190~\gev, \quad
160~\gev < m_{\tilde{b}_1} < 250~\gev .
\ee
As expected, $m_{\tilde{t}_1}$ and $m_{\tilde{b}_1}$ are strongly correlated, numerically shifted up by about 60 GeV. The difference between them is caused by the $D$-term and Yukawa contribution to the supersymmetric part of the stop/sbottom mass, as well as the mixing effects in the stop sector.
In our scenario with  $M_{\tilde{t}_R} \gg \sqrt{m_tA_t},~ M_{\tilde{Q}_3}$, the light stop mass  is approximately
\be
m_{\tilde{t}_1}^2 \approx  M_{\tilde{Q}_3}^2 + m_t^2 -
X_t^2 \frac{M_{\tilde{Q}_3}}{M_{\tilde{t}_R}} m_t^2, \ \ {\rm for}\ \
 X_t =\frac{A_t}{\sqrt{M_{\tilde{Q}_3} M_{\tilde{t}_R} } }.
\ee
From Fig.~\ref{fig:MA}(c), we see that $X_t$ is constrained in our scenario to be $3.5 < X_t < 4.5$. At the same time, $M_{\tilde{t}_R}$ lies between $1.5 - 2$ TeV.  Given that $M_{\tilde{Q}_3}$ is not very different from $m_t$ in our scenario, we see that the light stop mass is at most shifted by a fraction (about $30 \%$) of the top mass.
  We have also effectively decoupled $\tilde{b}_R$ by taking $M_{\tilde{b}_R}$ to be very heavy. Therefore, $m_{\tilde{b}_1} \sim M_{\tilde{Q}_3}$.
  The lightest sbottom would not be too much heavier than the lightest stop, as seen in Fig.~\ref{fig:M2}(a).
 A smaller $M_{\tilde{b}_R}$ could change the lower-lying spectrum of the sbottom sector, which will not be discussed in the current study.

\begin{figure}[t]
\begin{center}
\minigraph{7.5cm}{-0.15in}{(a)}{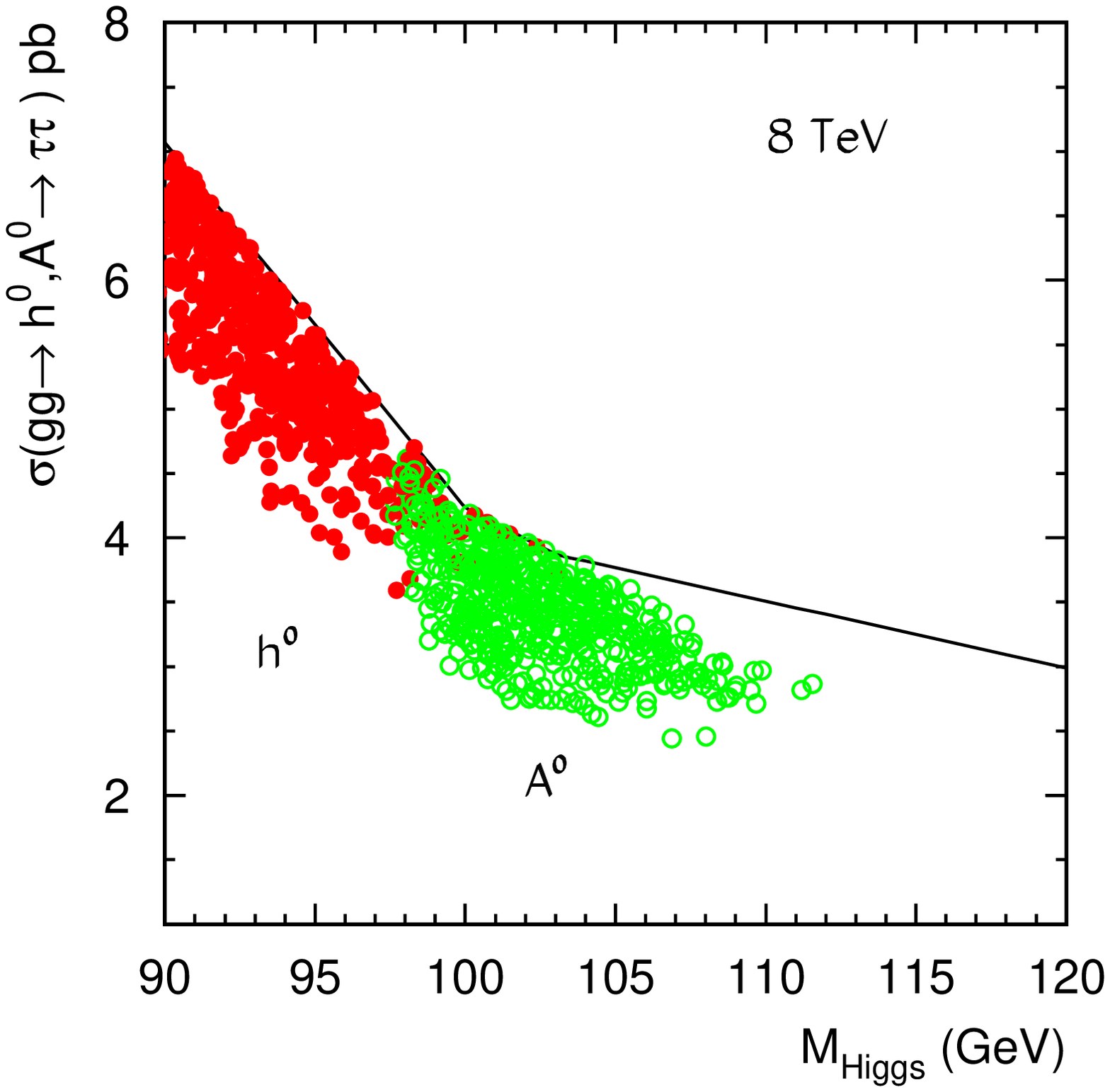}
\minigraph{7.5cm}{-0.15in}{(b)}{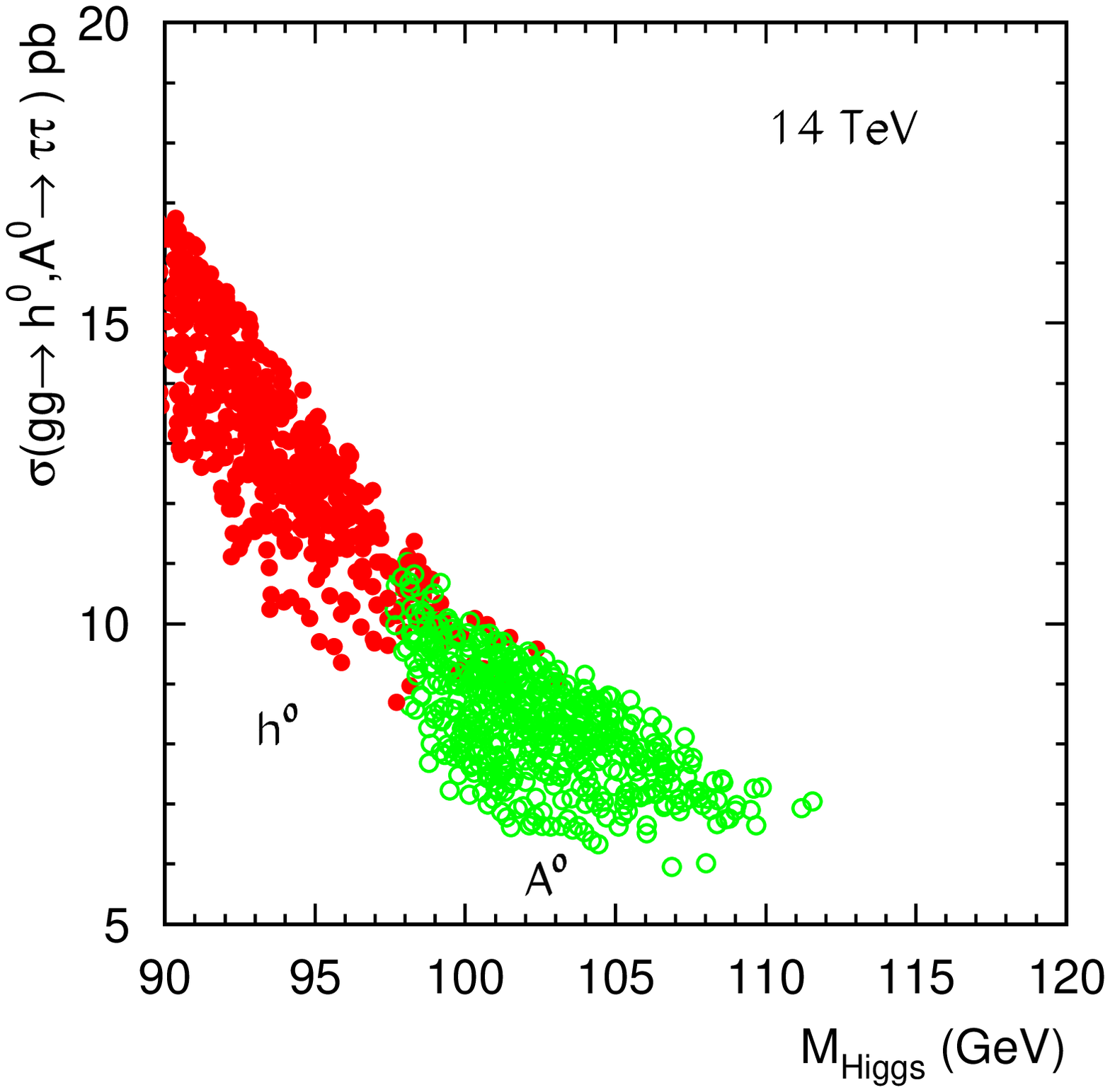}
\caption{Cross section for the process $\sigma(gg\to h^0,A^0\to \tau^+\tau^-)$ as a function of the corresponding non-SM Higgs mass at the LHC for (a) 8 TeV and (b) 14 TeV. The red dots (for $h^0$) and green dots (for $A^0$) are our surviving points. The reverse-engineered $\sigma\times{\rm Br}$ limit from CMS \cite{mssmtau} is also displayed by the solid line in (a).}
\label{fig:tautau_exclusion}
\end{center}
\end{figure}

Shown in Fig.~\ref{fig:M2}(b) is the allowed region for  $M_2$ versus $M_{\tilde{Q}_3}$. The plot exhibits certain correlation between $M_{2}-M_{\tilde{Q}_3}$, as already discussed in Eq.~(\ref{eq:bsg_a}).
The mass for the Wino, $M_2$, is largely constrained by $\bsg$ to be small as discussed earlier. We find it in the range
 \begin{equation}
 100 \gev < M_2 < 250 \gev.
 \end{equation}
Also shown in Figs.~\ref{fig:M2}(c) and (d) are relations among the physical masses: the Wino mass versus $m_{\tilde{t}_1}$ and $m_{\tilde{b}_1}$, respectively.

Given the stringent bound from the direct searches for $gg\to h^0/A^0\to \tau^+\tau^-$ at the LHC \cite{mssmtau} and in anticipation of the improvement in the near future, we check our solutions further against the CMS search limit as shown in Fig.~\ref{fig:tautau_exclusion}. We adopted the reverse-engineered limit based on the LHC direct search \cite{mssmtau} shown by the solid line in (a). The package HiggsBounds-4.0.0 \cite{Bechtle:2008jh} was used for the evaluation. The predicted cross section at the 14 TeV LHC is presented in (b). Imposing the latest ATLAS charged Higgs search results \cite{ATLAS_Hpm} at $3\sigma$ level, on the other hand, did not further reduce the parameter space.

As a final remark, we consider the lightest supersymmetric particle (LSP) with R-parity to be a viable candidate for the WIMP dark matter. While neutral Wino could be the LSP, the stop/sbottom being the LSP should be avoided. The Bino mass $M_1$, however, is unconstrained in the MSSM since Bino does not contribute much to either the Higgs sector, or the flavor observables. We thus prefer the Bino as the LSP, with $M_1<M_{2}, m_{\tilde{t}_1}, m_{\tilde{b}_1}$.\footnote{The values of $M_1$ and $M_2$ are approximately the same as the physical masses of $\tilde\chi^0_1$ and $\tilde \chi^+_1/\tilde \chi^0_2$ for relatively large value of $|\mu|$. We thus use them as symbols for the corresponding physical masses over the whole context, for simplicity.}  In addition, scenarios with Wino LSP are disfavored by the sbottom searches at the LHC, as explained below in Sec.~\ref{Disc}.

%%%%%%%%%%%%%%%%%%%%%%%%%%%%%%%%%%

\section{Light SUSY Spectrum at colliders}
\label{Disc}

As we have shown in the previous sections, flavor constraints not only requires the presence of rather light third generation of squarks and Winos, but also restrict them to a narrow region of parameter space. Therefore, our scenario predicts the distinct feature of collider signals from such a light SUSY spectrum.
In fact, current searches already put additional strong limits on our scenario. In this section, we describe the parameter space which is still allowed taken into account current stop and sbottom search limits and argue that further dedicated searches should be carried out at the LHC.

\begin{figure}[t]
\centering
\includegraphics[scale=1.2,width=12cm]{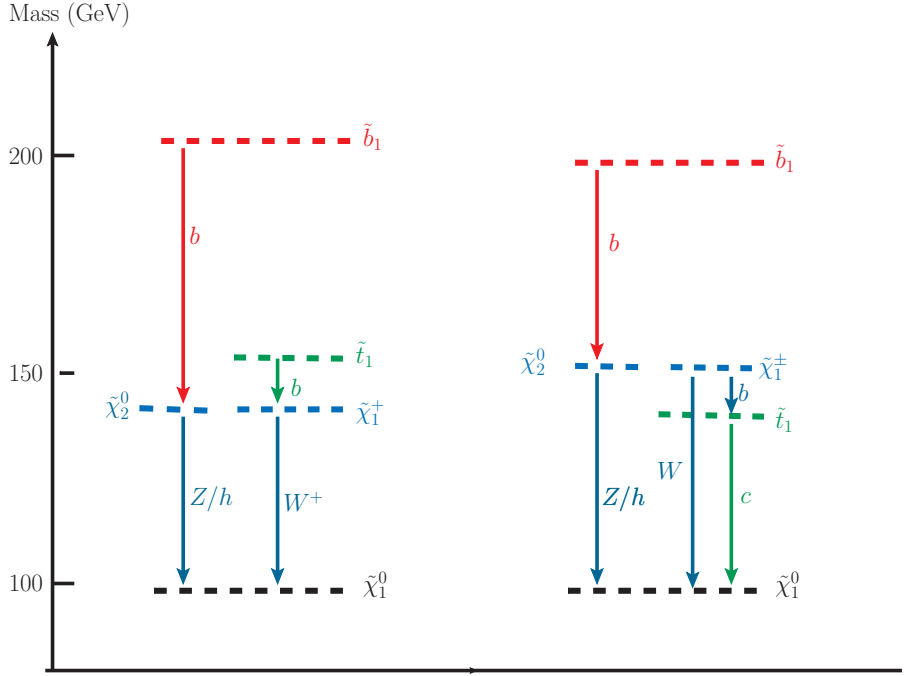}
\caption{Schematic drawing of two benchmark scenarios for Case A (left): $M_1<M_2<m_{\tilde{t}_1}<m_{\tilde{b}_1}$ and Case B (right): $M_1<m_{\tilde{t}_1}<M_2<m_{\tilde{b}_1}$.
The dominant decay modes are shown. The dashed horizontal lines are for illustrative purposes and do not represent precise mass values. }
\label{fig:benchmark}
\end{figure}

The relevant parameters under our consideration are $M_2,\ m_{\tilde{t}_1},\ m_{\tilde{b}_1}$, and $M_1$. The ordering of the Wino mass  with respect to the stop mass can be crucial in determining the pattern of stop decay, which in turn dictates the searching strategy. We thus categorize the allowed region into the following two cases:
\begin{itemize}
\item Case A: $M_1<M_2<m_{\tilde{t}_1}<m_{\tilde{b}_1}$;
\item Case B: $M_1<m_{\tilde{t}_1}<M_2<m_{\tilde{b}_1}$,
\end{itemize}
as illustrated in Fig.~\ref{fig:benchmark}.
The other possibility of $M_1<m_{\tilde{t}_1}<m_{\tilde{b}_1}<M_2$ is highly disfavored by sbottom search, as   explained below.

%%%%%%%%%%%%%%%%%%%%%

%%%%%%%%%%%%%%%%%%%%

\subsection{Stop/Sbottom Decay Patterns and Searches at the LHC}

\begin{figure}[t]
\centering
\includegraphics[width=0.49\textwidth]{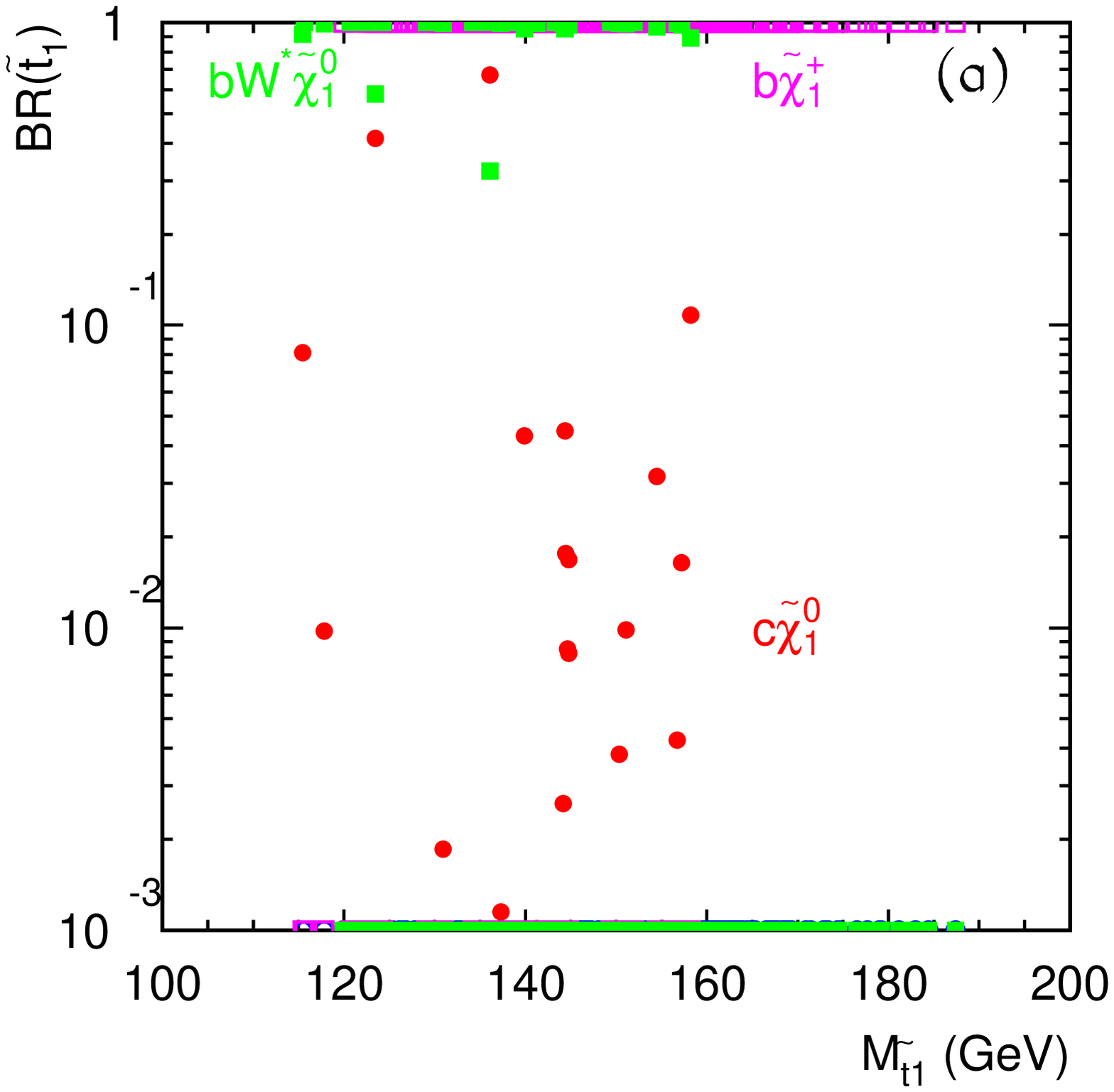}
\includegraphics[width=0.49\textwidth]{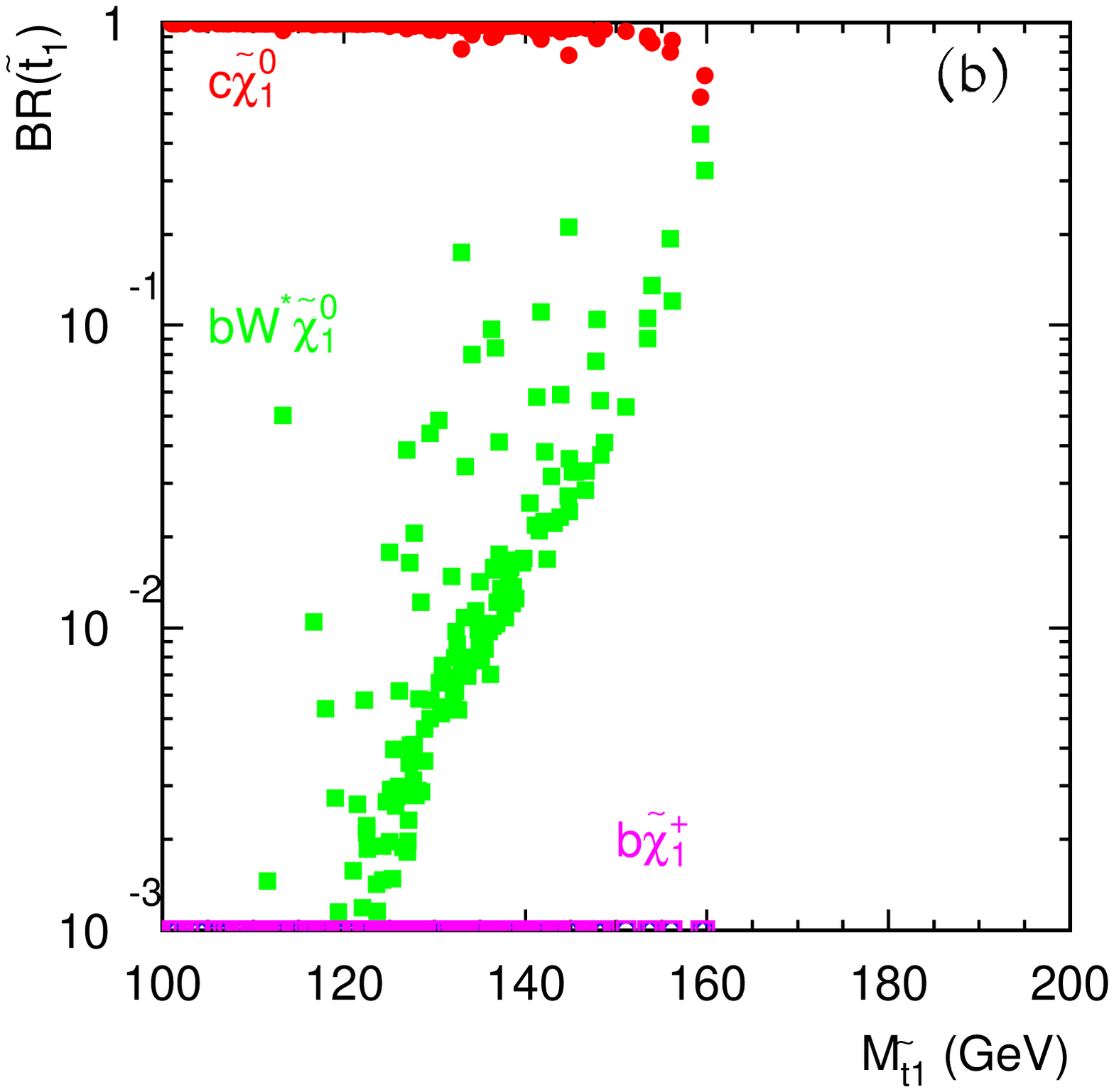}
\includegraphics[width=0.49\textwidth]{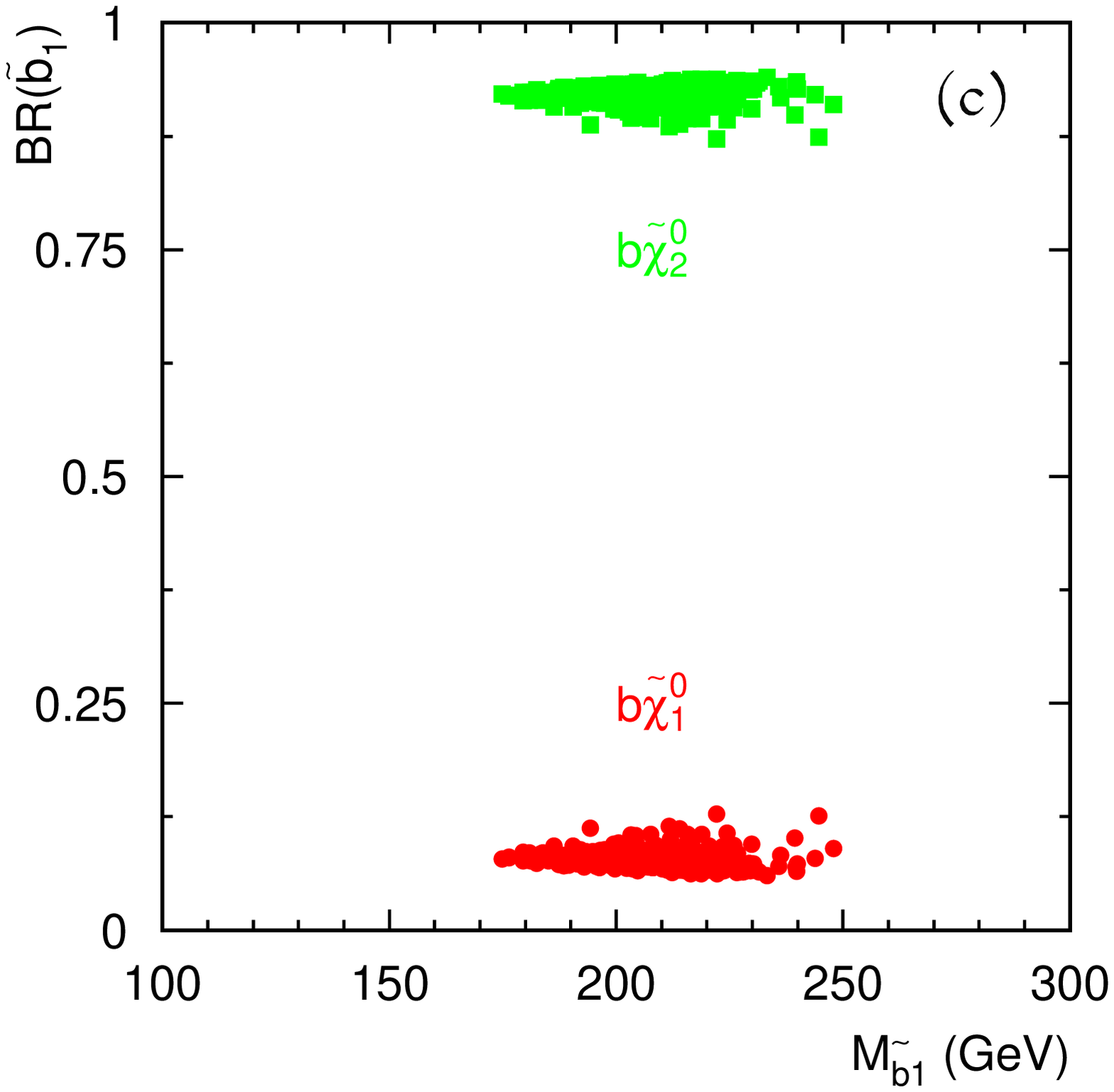}
\includegraphics[width=0.49\textwidth]{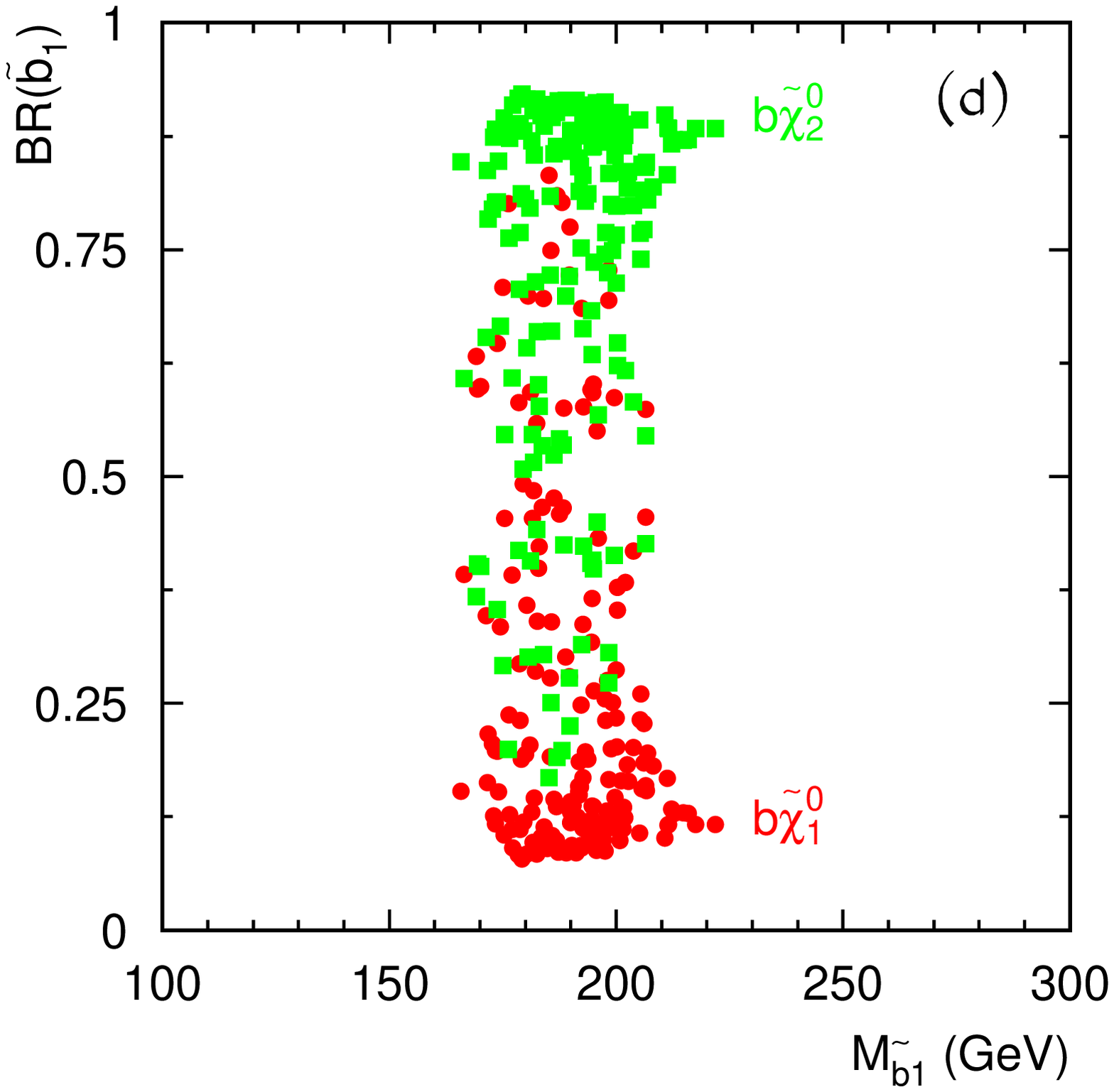}
\caption{Branching fractions of $\tilde{t}_1$ decay for (a) Case A: $M_1<M_2<m_{\tilde{t}_1}<m_{\tilde{b}_1}$ and (b) Case B: $M_1<m_{\tilde{t}_1}<M_2<m_{\tilde{b}_1}$, and  $\tilde{b}_1$ decay for (c) Case A and (d) Case B, respectively.}
\label{fig:BRs2-2}
\end{figure}

In Fig.~\ref{fig:BRs2-2}, we show the branching fractions of $\tilde{t}_1$ decay in panels (a), (b) and $\tilde{b}_1$ decay  in panels (c), (d) for Cases A and B, respectively.
We have kept $M_1=90$ GeV for illustration. A light stop dominantly decays via
\bea
&& \tilde t_1 \to b \tilde \chi^+_1 \ \ {\rm with} \ \
\tilde \chi^+_1 \to W^{+ (*)} \tilde \chi^0_1, \ \ {\rm for\ Case\ A},
\label{eq:stopdecA}
\\
&&
\tilde t_1 \to c \tilde \chi_1^0,  \qquad \qquad  \qquad \qquad  \qquad
{\rm for\ Case\ B}.
\label{eq:stopdecB}
\eea
If the Winos are lighter than the stop as in Case A, the dominant decay mode is as in Eq.~(\ref{eq:stopdecA}). There are dedicated LHC searches for this channel~\cite{ATLASstop,CMSstop}. However, the reaches of such searches are limited by the lepton and/or $b$-jet thresholds, governed by the SUSY particle mass differences. For example, for $m_{\tilde t_1} \sim 160$ GeV, a limit only exists for $m_{\tilde{\chi}_1^0} < 80 $ GeV.
In Case B where $M_2 > m_{\tilde t_1}$, the loop mediated FCNC decay mode $\tilde t_1 \to c \tilde \chi_1^0$ could become dominant, especially when $m_{\tilde t_1} - m_{\tilde{\chi}_1^0} < m_W $.
In such a case, the strongest experimental limit comes from the Tevatron~\cite{Tevatronstop}, which is only relevant for $m_{\tilde{\chi}_1^0} < 80$ GeV. Phenomenological studies~\cite{Krizka,Delgado:2012eu} have claimed that additional LHC analysis can be re-casted to set stronger limits on this and on the competing 4-body stop decay modes. These results,  if validated by the experimental collaborations, will further reduce the viable parameter region of our scenario.
On the other hand, if a stop signal is established at the LHC, the decay patterns as in Eqs.~(\ref{eq:stopdecA})
and (\ref{eq:stopdecB}) would help to determine the mass relation for the SUSY particles as Case A or B.

In parallel to the above discussion, the sbottom dominantly decays to one possible mode
\bea
\tilde{b}_1 \to b \tilde \chi^0_2 \ \ {\rm with}\ \
\tilde \chi^0_2 \to Z^{(*)} /h^{0(*)} \tilde{\chi}_1^0, \ \ {\rm for\ Cases\ A \ and \ B}.
\label{eq:sbdecA}
\eea
The other possibility that $\tilde{b}_1 \to b \tilde \chi^{0}_{1}$ dominates for the case of $M_1<m_{\tilde{t}_1}<m_{\tilde{b}_1}<M_2$ is highly disfavored by direct sbottom searches, as will be explained below.
Even though relatively suppressed by the available phase space, ${\rm BR}(\tilde{b}_1 \rightarrow b \tilde{\chi}_2^0)$ is typically around 80\% $-$ 90\%, given the larger $SU(2)_L$ coupling.
With the further decay of $\tilde{\chi}_2^0$ into $h^{0(*)} \tilde{\chi}_1^0$ or $Z^{(*)} \tilde{\chi}_1^0$,
these longer decay chains tend to give softer decay products and smaller missing energy as well. There are SUSY searches focusing on final states with higher multiplicity of jets~\cite{MET}. However, such searches typically require more energetic final states, for example the scalar sum of the transverse energy of jets $H_T > 600$ GeV. Therefore, they will not cover the low mass scenario we are considering. Sbottom decay also gives rise to multi-lepton signals, although they are suppressed by $Z$ leptonic branching ratio. A detailed recast of such limits on our scenario is beyond the scope of this paper.

Recently CMS has put the lower limit on the $m_{\tilde{\chi}_1^\pm, \tilde{\chi}_2^0}$ to 330 GeV with $m_{\tilde{\chi}_2^0}-m_{\tilde{\chi}_1^0}>m_Z$ and the assumption that  ${\rm BR}(\tilde\chi_2^0 \rightarrow Z \tilde\chi_1^0)={\rm BR}(\tilde\chi_1^\pm \rightarrow W^\pm \tilde\chi_1^0)=100$\%~\cite{cmsgaugino}. This limit would not directly constrain our  case with  smaller mass splitting $m_{\tilde{\chi}_2^0}-m_{\tilde{\chi}_1^0}$ and possible suppression of ${\rm BR}(\tilde\chi_2^0 \rightarrow Z \tilde\chi_1^0)$.

There are also strong limits from sbottom searches using the standard decay channel $\tilde{b}_1 \to b \tilde \chi^{0}_{1}$, even if the spectrum is somewhat squeezed~\cite{sbottom}.  ATLAS searches of $bb+\met$ final states with sbottom pair production excludes sbottom mass up to 650 GeV for $m_{\tilde{b}_1} - m_{\tilde{\chi}_1^0} \gtrsim$ 20 GeV~\cite{sbottom}, while the CMS limits of $m_{\tilde{b}_1} \gtrsim$ 650 GeV only applies to $m_{\tilde{b}_1}-m_{\tilde{\chi}_1^0} \gtrsim$ 200 GeV~\cite{sbottomcms}, with the assumption of ${\rm BR}(\tilde{b}_1 \to b \tilde \chi^{0}_{1})=100\%$. Since the sbottom is always somewhat heavier than the stop, it is typical for the mass splitting between the sbottom and the LSP to be sizable with  $m_{\tilde{\chi}_1^0} \leq m_{\tilde{t}_1}$. As a result, the case with only light wino LSP, which is the simplest possibility to satisfy the flavor constraints, has already been ruled out by direct searches at the LHC.   For the mass spectrum of $M_1<m_{\tilde{t}_1}<m_{\tilde{b}_1}<M_2$, the dominant decay mode for $\tilde{b}_1$ is also $\tilde{b}_1 \to b \tilde \chi^{0}_{1}$. In this case, ${\rm BR}(\tilde{b}_1 \to b \tilde \chi^{0}_{1})=100\% $, thus this scenario is   excluded by the sbottom searches as well. This exclusion pushes the upper limit of $M_2$ down to about 200 GeV as seen in Fig.~\ref{fig:M2}(d).
It is thus important to observe the gaugino mass limit as
 \begin{equation}
100 \gev < M_2 < 160 \gev\ {\rm for\ Case\ A},\ \ \  {\rm and}\ \ \
100 \gev < M_2 < 200 \gev\ {\rm for\ Case\ B}.
\label{eq:m2}
 \end{equation}
Viable points corresponding to Case (A) and (B) are indicated in Fig.~\ref{fig:M2} in red and green, respectively.

The conclusion of this section can be summarized with Fig.~\ref{fig:benchmark}, in which we show the mass hierarchies and dominant decay modes in the two benchmark cases.
We emphasize that the mass scales (as indicated by the horizontal bars) are only indicative of the size of viable values, and the bino mass can be raised or lowered without   significantly changing the flavor constraints.
A general lesson from the LHC bounds is that dedicated searches with rather soft decay products should be devised
when considering the light SUSY spectrum.

%%%%%%%%%%%%%%%%%%%%%%%%%%%%%

\subsection{Light SUSY spectrum and the ILC}

Although the super-partners in a light spectrum could be copiously produced at the LHC, identification of the signals can be very challenging in certain parameter region because of the hostile background environment.
As discussed in the previous session, for the same reason that light SUSY partners may have not been seen because of the soft products in the mass degenerate situation, this scenario may persist to prevent us from observing the SUSY signals. On the other hand, once crossing the mass threshold, the super-partners can be produced via the $SU(2)_L$ gauge interaction in $e^{+}e^{-}$ collisions
\be
e^{+}e^{-} \to
\tilde{t}_1 \tilde{t}_1^*, \quad
\tilde{b}_1 \tilde{b}_1^*, \quad {\rm and}\quad
\tilde{\chi}_1^+ \tilde{\chi}_1^-.
\ee
As long as the mass difference between the produced particle and the LSP $\tilde{\chi}_1^0$ is higher than the detection threshold for leptons and jets, typically about a few GeV, the signal identification should be straight forward, without suffering from much background.
For instance, at the ILC, even with the early phase with a C.M.~energy $\sqrt s = 250 -500$ GeV and integrated luminosity $250-500$ fb$^{-1}$ \cite{Brau:2012hv}, it would be essentially adequate to fully cover the low mass SUSY searches in the non-decoupling Higgs scenario.

%%%%%%%%%%%%%%%%%%%%%%%%%%%%%%%%%%%%%%

\section{Conclusions}
\label{Concl}

The non-decoupling scenario of the MSSM Higgs sector
presents an interesting extension beyond the SM and could be of immediate relevance for the LHC phenomenology after the discovery of the SM-like Higgs boson.

By scanning the SUSY parameter space and zooming in certain narrowly defined region, we demonstrated that this scenario is consistent with the current SUSY search limits. Moreover it can pass stringent flavor constraints from rare $B$ decay measurements, provided there are other light SUSY particles to contribute in the loop induced processes.
The regions of the parameter space which can realize such  cancellations are often missed by generic multiple parameter scans. Therefore, special caution is called for when scrutinizing the SUSY parameter space.

To satisfy the flavor constraints in the non-decoupling scenario, in particular from $\bsg$, it is necessary to have  light super-partners with quite specific properties, as summarized in Section \ref{ScanS}, Figs.~\ref{fig:MA} and \ref{fig:M2}. In particular, a (left-handed) stop, a (left-handed) sbottom, and Wino-like gauginos are required to be lighter than about 250 GeV. We also checked our solutions against the LHC Higgs searches \cite{mssmtau} $gg\to h^0/A^0\to \tau^{+}\tau^{-}$, and the results were plotted in Fig.~\ref{fig:tautau_exclusion}(a). Our prediction for the light non-SM Higgs boson $\tau\tau$ signal at the 14 TeV LHC was shown in Fig.~\ref{fig:tautau_exclusion}(b).
With respect to the latest results from ATLAS charged Higgs searches in the $\tau+$ jets channel \cite{ATLAS_Hpm}, we implemented their constraints on the branching fraction of $t \rightarrow b H^\pm$ at $3\sigma$ level of their errors. Should this constraint be further significantly tightened, say by a factor of few, then our scenario would not be adequate any longer. However, that might  in turn indicate the existence of more sophisticated underlying physics structure, such as modified $tbH^{\pm}$ coupling or reduced BR$(H^{-} \to \tau\nu$).

In Sec.~\ref{Disc}, we discussed the current limits from the relevant direct SUSY searches at the LHC and Tevatron, and identified two scenarios which are still allowed, as illustrated in Fig.~\ref{fig:benchmark}. The viable mass ranges of the light SUSY particles, their corresponding branching fractions and the typical signal rates at the LHC are summarized in Tables \ref{tab:summaryAB}. Although these light SUSY states may be copiously produced at the LHC
 as seen in the Table, the signal identification could still be challenging at the LHC due to the large SM backgrounds and disfavored kinematics.  An ILC with several hundred GeV center of mass energy could be more beneficial to discover and study those low mass states once crossing their mass threshold in a definitive way.

%%%%%%%%%%%%%%%%%%%%%%%

\begin{table}[t]
\begin{tabular}{|c|c|c|c|c|}
\hline
Case A   & mass range       & search mode BR &  LHC signal rate \\ \hline
$\tilde{\chi}_1^\pm$         & $100-160$ GeV & ${\rm BR}(\tilde{\chi}_1^\pm\to q\bar{q}'+ \tilde{\chi}_1^0)\approx 70\%$   & $\sigma(\tilde{\chi}_1^+\tilde{\chi}_1^-)\times {\rm BR}^2\approx 5-0.9$ pb                   \\ \hline
$\tilde{\chi}_2^0$          & $100-160$ GeV & ${\rm BR}(\tilde{\chi}_2^0\to b\bar{b}+ \tilde{\chi}_1^0)\approx 100\%$                 &        $\sigma(\tilde{\chi}_2^0\tilde{\chi}_1^\pm)\times {\rm BR}^2\approx 9.6-1.6$ pb           \\ \hline
$\tilde{t}_1$         & $110-190$ GeV & ${\rm BR}(\tilde{t}_1\to b\tilde{\chi}_1^+/b W^{+(\ast)}+\tilde{\chi}_1^0)\approx 100\%$                & $\sigma(\tilde{t}_1\tilde{t}_1^\ast)\times {\rm BR}^2\approx 10^3-10^2$ pb                   \\ \hline
$\tilde{b}_1$   & $170-250$ GeV & ${\rm BR}(\tilde{b}_1\to b+\tilde{\chi}_2^0)\approx 85\%$                 & $\sigma(\tilde{b}_1\tilde{b}_1^\ast)\times {\rm BR}^2\approx 110-22$ pb                   \\ \hline
Case B  &       &  & \\ \hline
$\tilde{\chi}_1^\pm$         & $110-200$ GeV & ${\rm BR}(\tilde{\chi}_1^\pm\to b+ \tilde{t}_1)\approx 100\%$   & $\sigma(\tilde{\chi}_1^+\tilde{\chi}_1^-)\times {\rm BR}^2\approx 7.6-0.85$ pb                   \\ \hline
$\tilde{\chi}_2^0$          & $110-200$ GeV & ${\rm BR}(\tilde{\chi}_2^0\to b\bar{b}+ \tilde{\chi}_1^0)\approx 100\%$                 &        $\sigma(\tilde{\chi}_2^0\tilde{\chi}_1^\pm)\times {\rm BR}^2\approx 9.6-0.96$ pb           \\ \hline
$\tilde{t}_1$         & $100-160$ GeV & ${\rm BR}(\tilde{t}_1\to c+\tilde{\chi}_1^0)\approx 100\%$                & $\sigma(\tilde{t}_1\tilde{t}_1^\ast)\times {\rm BR}^2\approx 10^3-200$ pb                   \\ \hline
$\tilde{b}_1$   & $170-220$ GeV & ${\rm BR}(\tilde{b}_1\to b+\tilde{\chi}_2^0)\approx 85\%$                 & $\sigma(\tilde{b}_1\tilde{b}_1^\ast)\times {\rm BR}^2\approx 110-36$ pb                   \\ \hline
\end{tabular}
\caption{Predicted light SUSY particles, their search modes and the signal rates for Cases A and B at the 14 TeV LHC.
}
\label{tab:summaryAB}
\end{table}

We also emphasize that these cases discussed here represent complementary opportunities to explore the non-decoupling scenario in addition to the direct searches for the Higgs bosons in the MSSM.

\acknowledgments
T.~L. and L.T.W. would like to thank Wolfgang Altmannshofer for helpful discussions.
The work of T.H.~was supported in part by the U.S.~Department of Energy under Grant No. DE-FG02-95ER40896 and in part by the PITT PACC. The work of T.L. is supported by the Australian Research Council. The work of S.S. is supported by the Department of Energy under Grant~DE-FG02-13ER41976.
L.T.W. is supported by the DOE Early Career Award under grant de-sc0003930. L.T.W was also supported in part by the Kavli Institute for Cosmological Physics at the University of Chicago through grant NSF PHY-1125897 and an endowment from the Kavli Foundation and its founder Fred Kavli.
T.H., S.S. and L.T.W. would also like to thank the Aspen Center for Physics for the hospitality during which part of this work was carried out. ACP is supported by NSF under grant 1066293.

%%%%%%%%%%%%%%%%%%%%%%%%%%%%%%%%%%%%%%%%%%%%
\appendix

\section{Formulae for loop functions relevant for $\bsg$. }
\label{sec:appendix_bsg}
\begin{eqnarray}
f_7(x) = {x\over 4}\left[{3-5x\over 6(x-1)^2}+{3x-2\over 3(x-1)^3}{\rm log} \ x\right],\quad
f_8(x) ={x\over 4}\left[{3-x\over 2(x-1)^2}+{-1\over (x-1)^3}{\rm log} \ x\right].
\label{eq:f78}
\end{eqnarray}
\begin{eqnarray}
g^{(a)}_7(x,y)&=&{2x-3\over 6(x-y)(x-1)^3}{\rm log} \ x+{2y-3\over 6(x-y)(1-y)^3}{\rm log} \ y+{-7x-7y+5xy+9\over 12(x-1)^2(y-1)^2},
\label{eq:g7a}\\
g^{(a)}_8(x,y)&=&{x\over 2(x-y)(x-1)^3}{\rm log} \ x+{y\over 2(x-y)(1-y)^3}{\rm log} \ y+{x+y+xy-3\over 4(x-1)^2(y-1)^2}.
\label{eq:g8a}
\end{eqnarray}
\begin{eqnarray}
g^{(b)}_7(x,y)&=&
{y(3-2x)\over 3(y-x)(1-x)^3}{\rm log} \ x+{y^2(-3+2y)\over 3x(x-y)(1-y)^3}{\rm log} \ y+{y(-7+5x+5y-3xy)\over 6x(1-x)^2(1-y)^2},\label{eq:g7b}\\
g^{(b)}_8(x,y)&=&{xy\over (x-y)(1-x)^3}{\rm log} \ x-{y^3\over x(x-y)(1-y)^3}{\rm log} \ y+{y(1+x+y-3xy)\over 2x(1-x)^2(1-y)^2}.
\label{eq:g8b}
\end{eqnarray}
\begin{eqnarray}
g^{(c)}_7(x)={7x^2-5x-8\over 72(x-1)^3}+{x(3-2x)\over 12(x-1)^4}{\rm log} \ x,\ \ \
%-{2+11x-7x^2\over 18(1-x)^4}-{x(3-x^2)\over 6(1-x)^5}{\rm log} \ x.
g^{(c)}_8(x)={2x^2+5x-1\over 24(x-1)^3}+{-x^2\over 4(x-1)^4}{\rm log} \ x.
\label{eq:g78c}
\end{eqnarray}
\begin{eqnarray}
g^{(e)}_7(x,y)&=&{x(2x-3)\over 6(x-y)(x-1)^3}{\rm log} \ x+{y(2y-3)\over 6(x-y)(1-y)^3}{\rm log} \ y+{-5x-5y+3xy+7\over 12(x-1)^2(y-1)^2},\label{eq:g7e}\\
g^{(e)}_8(x,y)&=&{x^2\over 2(x-y)(x-1)^3}{\rm log} \ x+{y^2\over 2(x-y)(1-y)^3}{\rm log} \ y+{-x-y+3xy-1\over 4(x-1)^2(y-1)^2}.
\label{eq:g8e}
\end{eqnarray}

%%%%%%%%%%%%%%%%%%%%%%%%%%%%%%%%%%%%%%%%%%%%%%%%%%%%%%%%%%%%%%%%%%%%
%\input{reference.tex}

\end{document}